# On the Role of Color Temperature and Color Rendering Index of White-Light LEDs on the Theoretical Efficiency Limit of Indoor Photovoltaics

Aditi Sharma[1], Ian Yum[1], Alexander A. Guaman[1], Gavin A. Akmam[1], Jason A. Röhr[1,*]

[1]General Engineering, Tandon School of Engineering, New York University, Brooklyn, 11201 New York, United States of America

*jasonrohr@nyu.edu

## ABSTRACT

As the Internet of Things (IoT) continues to grow, the demand for sustainable indoor power sources is increasing. Indoor photovoltaics (IPVs), which are currently in development, present a renewable solution but must be designed to match specific light sources to maximize efficiency. While previous studies have emphasized the role of white-light LED correlated color temperature (CCT) in determining IPV efficiency and optimum bandgap energy, the role of color rendering index (CRI) remains less understood. In this study, we employ detailed-balance calculations to assess the theoretical maximum efficiency and optimal bandgap energies of IPVs under commercial white-light LED illumination varying in both CCT (2200 K to 6500 K) and CRI (70, 80, and 90). Our results confirm that lower CCTs yield higher efficiencies and lower optimal bandgaps; however, contrary to prior assumptions that CRI has negligible impact on IPV material choice and performance, we demonstrate that commercial high-CRI LEDs, where said high CRI was achieved by extending the emission into the red, can necessitate the use of materials with significantly lower bandgap energies for optimum efficiency. We also evaluate the performance of various IPVs at fixed bandgaps, revealing that while optimal IPV performance is achieved with wide-bandgap materials under lower CRI lighting, mature technologies like silicon and CdTe can benefit from high-CRI illumination. These findings underscore the need to consider color characteristics beyond CCT when evaluating solar cells for indoor IoT applications.

## INTRODUCTION

The goal of the IoT is to create a network of interconnected devices that can collect and exchange data without human intervention, thereby automating tasks and improving decision-making via data analysis. The ultimate hope is to enhance efficiency and productivity across various industries and aspects of daily life. However, wide-spread implementation of the IoT faces a challenge in sustainably powering a rapidly growing number of devices—with an estimated 40 billion integrated devices by 2027.[1,2] While many IoT devices currently consume only minimal power, ranging from nW to mW, the explosive growth of IoT technologies underscores the need for utilizing sustainable energy sources. Indoor photovoltaics (IPVs) provide a promising, renewable alternative, offering the potential to reduce the environmental impact associated with battery replacement and disposal.[1–3] While IPVs have powered rudimentary indoor devices for decades, for example pocket calculators, and even though there is a drive towards reducing the power needed for IoT infrastructure, the growing number of devices are calling for efficient IPVs to be developed. Moreover, as emission profiles vary between indoor light sources, including those of commercial white-light LEDs, and since the choice of light source determines what material should be chosen as the photo-active component, there is not only a need to develop efficient IPVs, but also a need to understand how to optimize IPVs to match specific indoor lighting environments.

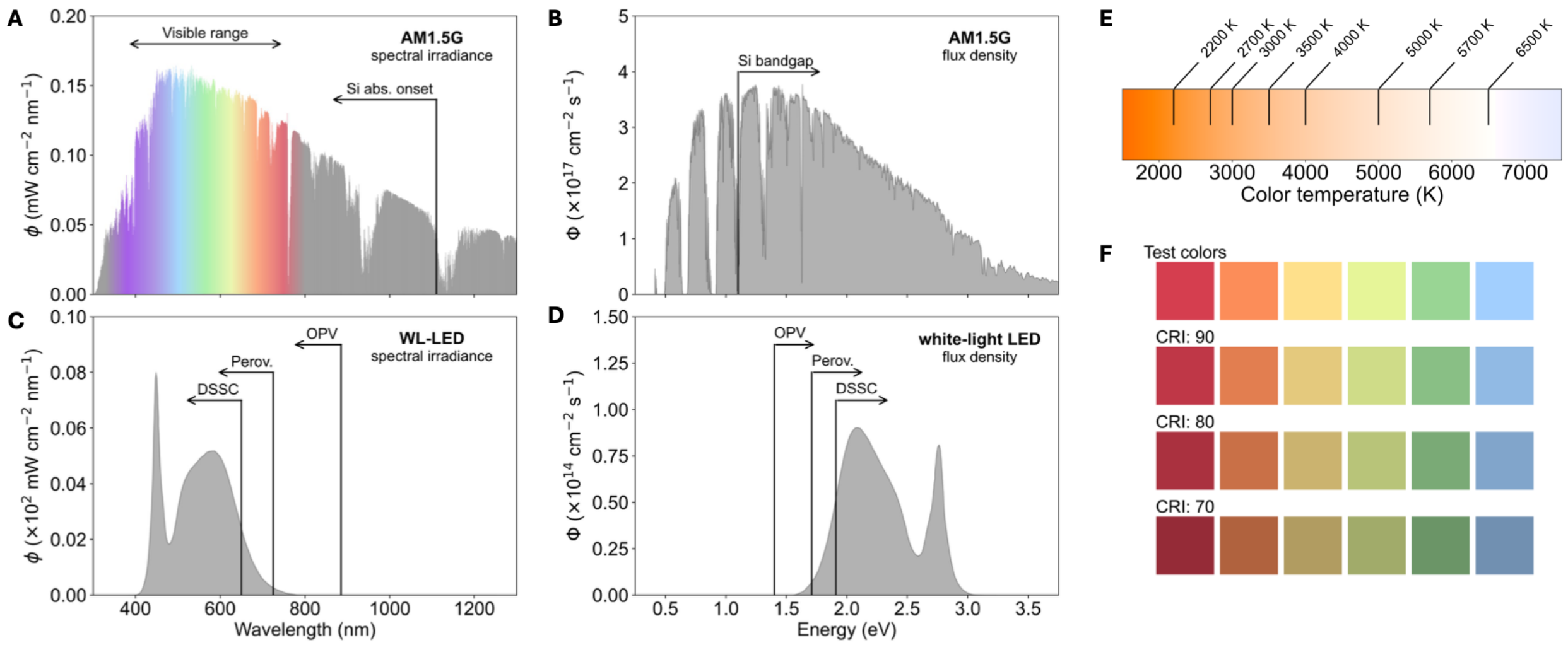

**Figure 1 – A,** AM1.5G solar irradiance, indicating the visible range of the solar spectrum and the absorption onset for crystalline silicon. **B,** Corresponding photon flux density of the AM1.5G spectrum and bandgap edge of crystalline silicon. **C,** Irradiance ($a = 0.1$) of a typical white-light LED (here a Cree J Series LED with CCT: 5000 K and CRI: 80) and absorption onsets of DSSC, perovskite, and OPV materials used in record-efficiency IPVs. **D,** Corresponding photon flux density and bandgap values for DSSC, perovskites, and OPV materials. **E,** Visualization of CCT, highlighting the specific values of the employed LEDs. **F,** Visualization of CRI, showing how certain test colors can vary with decreasing CRI.

The spectral irradiance spectrum of the Sun spans a wide range of wavelengths, from ultraviolet (UV) to the infrared (IR), with most of its irradiance concentrated in the visible and near-IR regions (**Fig. 1A,B**). Crystalline silicon (c-Si) solar cells, with a bandgap of ~1.1 eV (**Fig. 1B**), corresponding to an absorption onset at a wavelength of approximately ~1110 nm (**Fig. 1A**), are efficient at converting sunlight into electricity as its (albeit indirect) bandgap is near the theoretical optimum.[4,5] This, along with the abundance of Si and its well-established manufacturing processes, makes it a leading material for outdoor photovoltaic applications. In contrast, indoor lighting, such as white-light LEDs, not only emits at lower intensities but also within a much narrower spectral range, typically solely in the visible (**Fig. 1C,D**). This shifts the calculated theoretical maximum to much larger values of ~1.8 to ~2.0 eV.[6,7] As c-Si solar cells rely heavily on absorption in the IR (**Fig. 1A,B**), they are therefore less efficient in typical indoor environments. Researchers are for that reason actively exploring alternative technologies for powering indoor IoT applications. Perovskite,[8,9] organic,[10,11] and dye-sensitized solar cells[12,13] have optical gaps (and therefore absorption onsets) that can be chosen to match emission profiles of indoor light sources (**Fig. 1C,D**), and have shown great promise for IPV applications with measured indoor power-conversion efficiencies ranging from 30 to 45%. Despite these successes, IPV efficiencies are still far from their calculated theoretical maxima.[6]

Laying the groundwork for understanding the theoretical efficiency limits and optimum bandgap values for IPVs, Freunek *et al*. used detailed-balance calculations, employing irradiance spectra from a set of common indoor light sources.[6] They found that the device efficiency was, unsurprisingly, highly dependent on how well the bandgap of the active material matches the emission spectrum of the light sources. In fact, they observed that the optimum bandgap for solar cells illuminated by white-light LEDs is in the range of 1.9 to 2.0 eV, with theoretical efficiencies approaching 60%. White-light LED emission spectra are defined by certain color characteristics, such as their correlated color temperature (CCT) and color rendering index (CRI). CCT refers to the perceived warmth or coolness of a given light source (measured in K), where a lower temperature indicates a warmer, red light and a higher temperature indicates a cooler, blue light (**Fig. 1E**). CRI is a measure of how accurately a given light source displays the perceived colors of an object, in comparison to a reference source, when illuminating a set of standardized colors (**Fig. 1F**). CRI scales from 0 to 100, with higher values corresponding to smaller differences between the light source and the light reference.

A greater CRI often comes at a higher price point and is achieved by either incorporating additional mono-chromatic LEDs into the device or by including a white-light emitting phosphor layer. While emphasis is often put on the CCT of the light source within the IPV literature, especially within the context of IPV characterization, that is typically not the case for CRI. Ho *et al.* used a similar approach to that of Freunek *et al.*, revealing, among other things, the significance of CCT in influencing IPV performance; colder, blue light generally results in lower theoretical efficiency at higher bandgap values.[7] Using an approach of forming white-light emission from a superposition of monochromatic LEDs, Zheng *et al*. found that while CCT of white-light LEDs indeed is a dominant factor controlling IPV efficiency, with the power conversion efficiency decreasing as color temperature increases, they observed that CRI affects IPV performance to a lesser extent regardless of the CCT values.[14]

Building on these past reports, we here investigate the theoretical maximum efficiencies and optimum bandgaps of IPVs modeled as if illuminated by a range of commercial white-light LEDs with varying CCT (2200 K to 6500 K) and CRI (70, 80, and 90), where the 90 CRI had expanded emission profiles into the red. Our analysis confirms that both the maximum PCE and optimum bandgap value of the solar cell varies with CCT of the LEDs. Contrary to previous reports, our calculations also reveal that the choice of LED CRI can be significant, with IPVs illuminated by the investigated 90 CRI white-light LEDs having slightly less maximum theoretical efficiency at significantly lower optimal bandgap values across all investigated CCT values due to the broader emission spectral profile into the red. However, we show that technologies such as c-Si and CdTe can benefit from illumination with such high-CRI light. This indicates the need for device engineers to account for a broader range of LED characteristics, beyond CCT alone, when designing and characterizing IPVs.

## METHODS

We utilized a detailed-balance model to calculate the maximum power-conversion efficiency of IPVs when illuminated by a range of arbitrarily chosen commercial white-light LEDs with varying CCT and CRI values. We obtained the LED spectral power distribution (SPD) spectra as a function of wavelength, $\varphi_0(\lambda)$, directly from online data sheets provided by Samsung (LM301Z+ and LH502D), Cree (X Lamp and J Series), and Lumileds (LUXEON 3030 HE and LUXEON 5050). Mean values and standard deviation of the LED SPD spectra at 4000 K, 5000 K, and 6500 K are shown for CRI values of 70, 80, and 90 in **Fig. 2A-C**, respectively. SPD spectra of all the individual LEDs investigated herein are shown in the Supplementary Material (Fig. S1-3). Out of the 121 spectra that were used within our calculations, very little difference was observed for LEDs sharing CCT and CRI characteristics, as shown by the relatively low standard deviation (**Fig. 2A-C**). Finally, not all LED variants contained the entire CCT range. As such, we omitted the 90 CRI 2200 K spectra from our analysis.

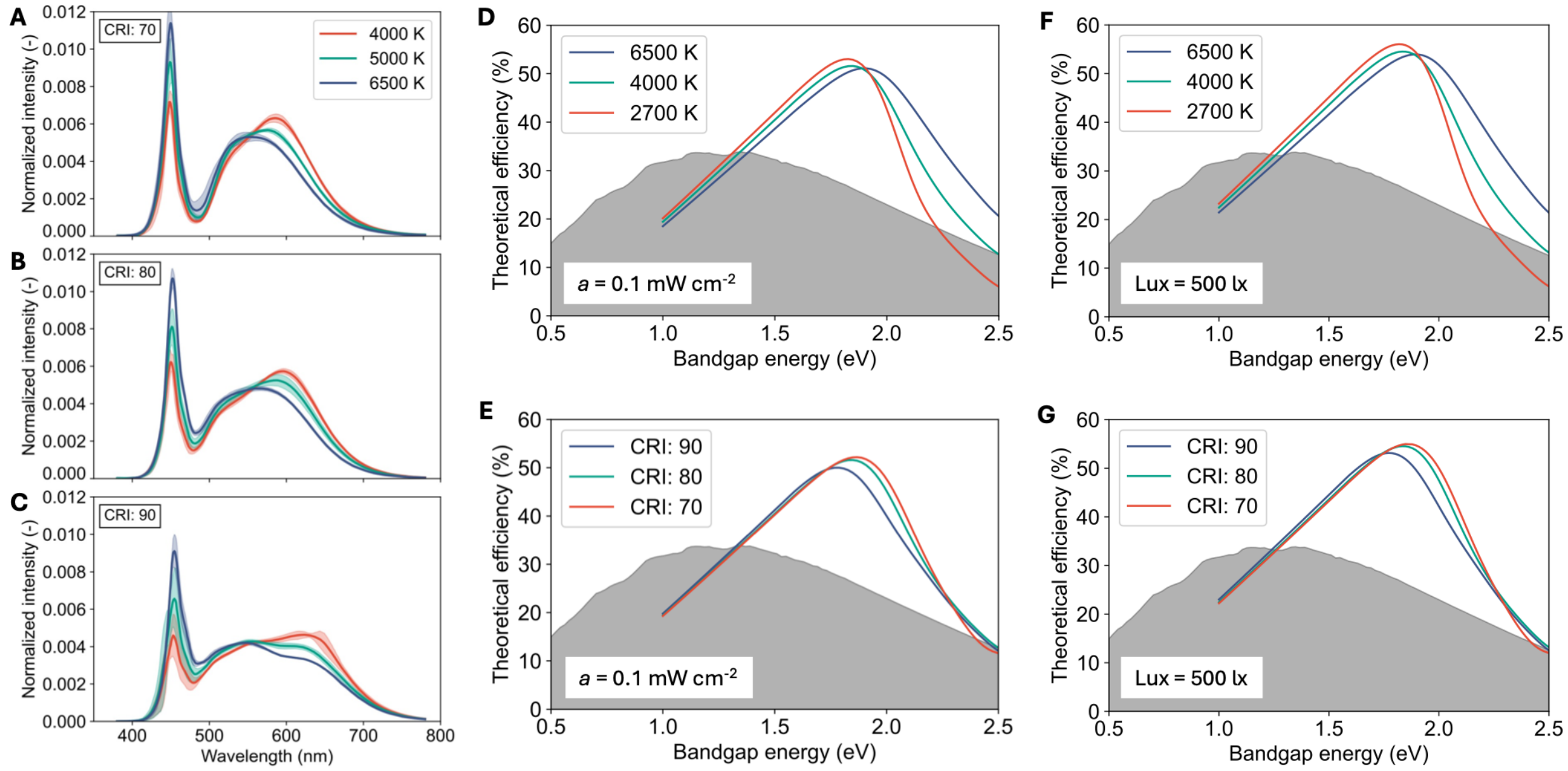

**Figure 2 –** Averaged, normalized emission spectra of white-light LEDs with CCT of 4000 K, 5000 K, and 6500 K for different value of CRI: **A,** 70, **B,** 80, and **C,** 90. **D,** Maximum theoretical efficiency as a function of bandgap energy for a solar cell illuminated under either standard AM1.5G conditions (gray) or illuminated by a LEDs under fixed irradiance ($a$ = 0.1 mW cm$^{-2}$) at different CCT values and a CRI of 80. **E,** Maximum theoretical efficiency of a solar cell illuminated by LEDs with different CRI values and CCT of 4000 K. **F,G,** Equivalent calculations, but under a fixed Lux of 500 lx.

Comparing IPVs illuminated by different white-light LEDs should be characterized on the basis of constant illuminance in Lux (lx);[2] however, to offer an easy relationship between power output and efficiency of the IPVs, we also compare LEDs with constant irradiance (mW cm$^{-2}$). To do this, each LED SPD spectrum was first normalized to have an area of unity,

$$\varphi_{\text{norm}}(\lambda) = \frac{\varphi_0(\lambda)}{\int_{\lambda_1}^{\lambda_2} \varphi_0(\lambda)\, d\lambda} \tag{1}$$

which allows us to set the irradiance of each LED to a set magnitude,

$$\varphi(\lambda) = a\varphi_{\text{norm}}(\lambda). \tag{2}$$

via the quantity, $a$, which has dimensions of mW cm$^{-2}$. This ensures that our comparisons isolate the spectral shape (CRI and CCT) at fixed and comparable total irradiance. White-light LEDs used for indoor lighting typically emit light at low irradiance (from ~0.1 mW cm$^{-2}$ to ~10 mW cm$^{-2}$) as compared to sunlight (100 mW cm$^{-2}$ for AM1.5G). For the present study, values for $a$ that match that of typical irradiance of indoor white-light LEDs were therefore chosen, i.e., $a = 0.1$ to 10 mW cm$^{-2}$.

We can now also set a fixed illuminance, $E$, for each LED via,

$$E = \kappa \int_{\lambda_1}^{\lambda_2} \varphi(\lambda)\, V(\lambda)\, d\lambda \tag{3}$$

where $\kappa$ is the luminous efficacy ($\kappa \sim 683$ lm W$^{-1}$) and $V$ is the CIE 1931 photopic eye response (as shown in the Supplementary Material), which defines the vision of the human eye under well-lit conditions. Within our model, the exercise now becomes solving for $a$ for a given target illuminance,

$$a = \frac{E^{\text{target}}}{\kappa \int_{\lambda_1}^{\lambda_2} \varphi_{\text{norm}} V(\lambda)\, d\lambda} \tag{4}$$

which in return set the magnitude of the irradiance of the white-light LED for a given illuminance, as defined by **Eq. 2**. Similar to how the irradiance values were chosen, we chose a range of illuminance values matching that of typical white-light LEDs, i.e., $E =$ 300 lx to 1000 lx.

The efficiency of a solar cell is determined by the ratio of output to input power (or power density),

$$\eta = \frac{p_{\mathrm{out}}}{p_{\mathrm{in}}}. \tag{5}$$

While $p_{\mathrm{out}}$ will be determined from the detailed-balance model, $p_{\mathrm{in}}$ for a given LED spectrum can be determined by integrating the chosen irradiance spectrum across all relevant wavelengths,

$$p_{\mathrm{in}} = \int_{\lambda_1}^{\lambda_2} \varphi(\lambda)\, d\lambda\,. \tag{6}$$

To calculate $p_{\mathrm{out}}$, we begin by converting the spectral irradiance into photon flux density, $\Phi(E)$ via a Jacobian transformation,

$$\Phi(E) = \varphi(\lambda)\frac{hc}{E^2} \tag{7}$$

where $h$ is Planck's constant, $c$ is the speed of light in vacuum, and $E$ is energy. If each photon with energy exceeding the bandgap, $E_{\mathrm{g}}$, of the active semiconductor contributes to a single electron-hole pair, the photocurrent density can be defined via the photon flux density as,

$$J_{\mathrm{ph}}(E_{\mathrm{g}}) = q\int_{E_{\mathrm{g}}}^{E_2} \frac{\Phi(E)}{E}\, dE \tag{8}$$

where $q$ is the elementary charge. Radiative recombination is inevitable for any warm body. Assuming ideal diode behavior, and that the quasi-Fermi-level splitting in the semiconductor is equal to the external voltage, $V$, the lowest possible radiative recombination current density is defined as,

$$J_{\mathrm{rec}}(V, E_{\mathrm{g}}, T_{\mathrm{c}}) = \frac{2\pi q}{h^3 c^2}\int_{E_{\mathrm{g}}}^{E_2} \frac{E^2}{\exp\left(\frac{E - qV}{k_{\mathrm{B}}T_{\mathrm{c}}}\right) - 1}\, dE \tag{9}$$

where $T_c$ is the solar cell temperature (which is assumed to be 300 K for all calculations herein). Finally, the total current density, $J_{\mathrm{tot}}$, is due to the difference between the photocurrent and the recombination current,

$$J_{\text{tot}}(V, E_{\text{g}}, T_{\text{c}}) = J_{\text{ph}}(E_{\text{g}}) - J_{\text{rec}}(V, E_{\text{g}}, T_{\text{c}}) \quad (10)$$

and $p_{\text{out}}$ is now given by the maximum of the product of the current density of the illuminated cell and the applied voltage, or equivalently, from the product of the short-circuit current density, $J_{\text{SC}}$, the open-circuit voltage, $V_{\text{OC}}$, and the fill factor, FF,

$$p_{\text{out}} = \max(|J_{\text{tot}}V|) = J_{\text{SC}}V_{\text{OC}}\text{FF} \quad (11)$$

Comparisons between the detailed-balance limit assuming standard AM1.5G illumination (the well-known Shockley-Queisser limit) and equivalent calculations using indoor LEDs are shown in **Fig. 2D,E** (for a fixed total irradiance of 0.1 mW cm$^{-2}$) and **Fig. 2F,G** (for fixed illuminance of 500 lx), highlighting that the maximum efficiency for IPVs exceeds that of outdoor PVs and is achieved with much larger bandgap values for the active semiconductor.

## RESULTS & DISCUSSION

The mean calculated theoretical maximum efficiencies and optimal bandgap energies, of all the investigated LEDs at three different fixed values of illuminance, are shown in **Fig. 3A-F**. Figures for equivalent calculations at fixed irradiance are shown in the Supplementary Material (**Fig. S4**), highlighting a similar trend as what was observed at fixed illuminance. The observed standard deviations within these calculations arise from the deviations in the SPD spectra between the various investigated LEDs (**Fig. 2A-C**). Similarly, because measured CCT values can deviate from values reported by manufacturers, we calculated the CCT values from the SPD spectra and we show their mean values and standard deviation for each CCT rather than the nominal values (see Supplementary Material for calculation details).

As reported elsewhere, IPVs illuminated by LEDs with lower CCT yield the highest theoretical maximum power-conversion efficiencies at the lowest optimal bandgaps.[7,14] The same trend is observed herein for all investigated LEDs regardless of the magnitude of the illuminance (**Fig. 3A,C,E**). Although optimal bandgap values remain relatively unchanged across illuminance levels, calculated efficiencies increase with illuminance, ranging from slightly above 51% (CRI: 90, CCT: 6500 K) to just above to 55% (CRI: 70, CCT: 2000 K) for $E = 300$ lx (**Fig. 3A**) and from just above 52% (CRI: 90, CCT: 6500 K) to close to 57% (CRI: 70, CCT: 2000 K) for $E = 1000$ lx (**Fig. 3E**). This increase in efficiency can be attributed to non-linear increases in $V_{\text{OC}}$ and FF with increased irradiance (**Fig. S5**) and therefore also illuminance (**Fig. S6**), with the predominant effect being a logarithmic increase in $V_{\text{OC}}$. Firstly, as radiative recombination, and therefore $J_{\text{rec}}$ is constant at a given $T_{\text{C}}$ (**Eq. 9**), a lower FF is expected at low irradiance due to a "*rounder*" *J-V* relationship. $J_{\text{ph}}$ increases linearly with irradiance (**Eqs. 7 & 8**); the $J_{\text{SC}}$ therefore also increases linearly (**Fig. S5A**). At higher irradiance, photocurrent dominates over the recombination current (**Eq. 10**), and the *J*-*V* curve becomes "*squarer*", increasing the FF (**Fig. S5B**). Secondly, $V_{OC}$ is logarithmically proportional to $J_{SC}$ ($V_{OC} \propto \ln J_{SC}$). $p_{out}$, which can be described by the product of $J_{SC}$, $V_{OC}$, and FF (**Eq. 11**), therefore increases superlinearly with increased irradiance (**Fig. S5C**), increasing the efficiency superlinearly as well (**Fig. S5D**).

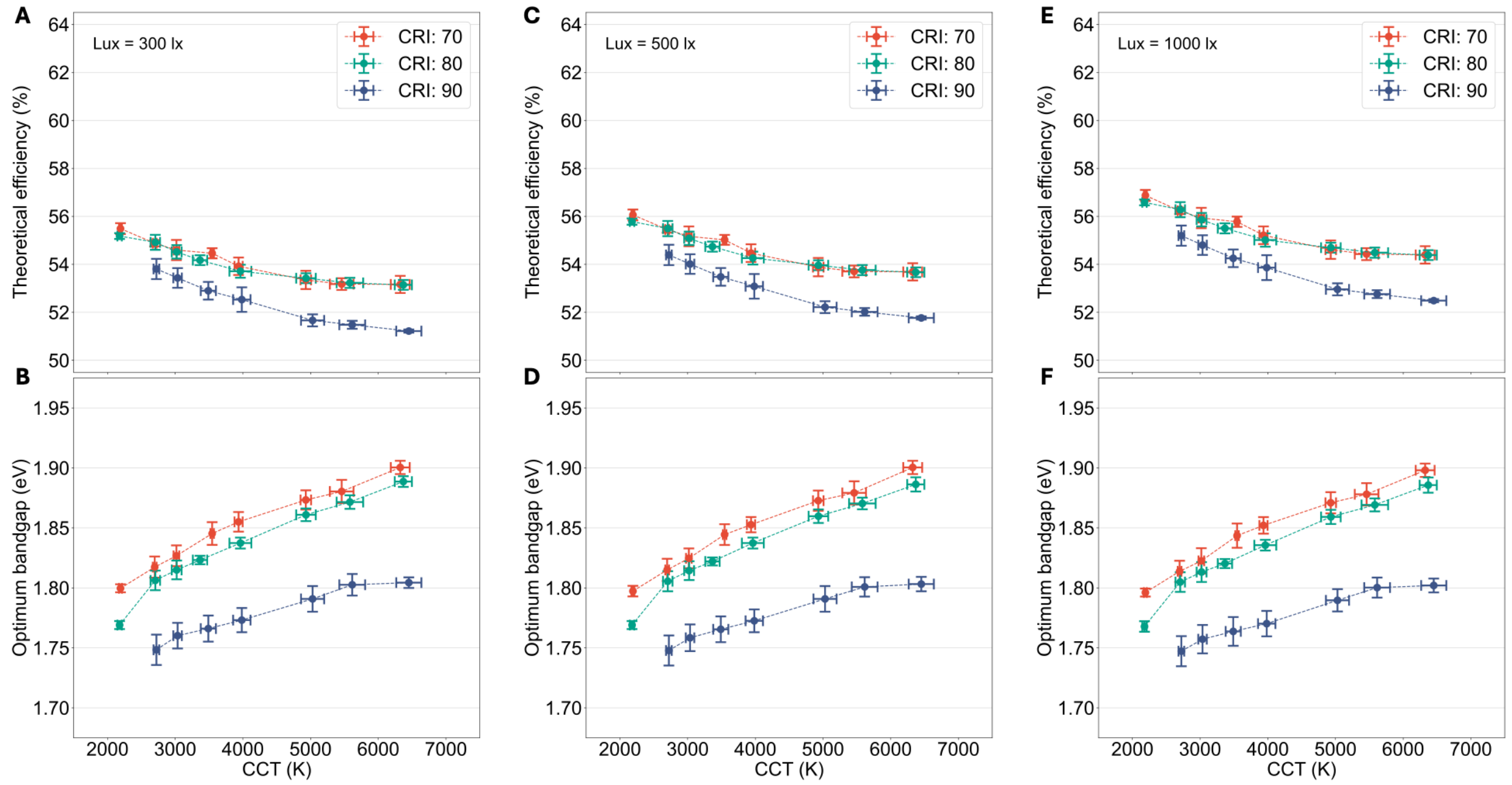

**Figure 3 - A,** Averages and standard deviations of calculated theoretical maximum efficiencies of IPVs illuminated by white-light LEDs (Lux = 300 lx) with different CCT values (ranging from 2200 K to 6500 K) and CRI values (70, 80, and 90). **B,** Corresponding averages of calculated optimum bandgap energies resulting in maximum theoretical efficiency. **C,D,** Similar calculations but for 500 lx. **E, F,** Calculations for 1000 lx.

Even though additional emissive components are found in the cyan-green part of the 80 CRI spectra as compared to 70 CRI (Fig. 2A,B) no significant differences are observed in their calculated results, regardless of the magnitude of the illuminance (**Fig. 3A-F**) or irradiance (**Fig. S4**), echoing the previous reports that the choice of CRI of the LEDs do not have a significant impact on the IPVs they are illuminating. However, a deviation is observed for IPVs illuminated by an LED with CRI of 90 yielding not only slightly lower efficiencies (close to 54% at 2700 K to ~51% at 6500 K for 300 lx), but yielding said efficiencies at significantly lower bandgap energies (~1.75 to ~1.80 eV as compared to ~1.76 eV to ~1.90 eV for 70 and 80 CRI LEDs). This is not entirely surprising, as the LEDs with CRI of 90 that we are investigating not only contain additional single-color emissive components but also broader emission profiles, with higher emission intensities at larger wavelengths as compared to the investigated spectra with CRI of 70 and 80 (**Fig. S7**). So, while it was previously reported that CRI can have a lesser impact on solar cell device efficiency,[14] we here show that if high CRI is achieved by spectral broadening into the red, lower bandgap materials should likely be utilized for optimal performance.

Finally, we investigate whether certain types of solar cells, spanning both mature and next-generation cells, can be affected by the choice of LED CCT and CRI. To do this, we calculate and compare the maximum theoretical efficiency at set bandgap values (**Fig. 4A-D**): 1.1 eV (corresponding to c-Si), 1.3 eV (corresponding to GaAs) 1.5 eV (corresponding to CIGS, CdTe, and record perovskite cells), and 1.8 eV (corresponding to a-Si:H and various next-generation cells). Equivalent calculations were performed at fixed irradiance, $a = 0.1$ mW cm$^{-2}$ (**Fig. S8**), showing a similar trend.

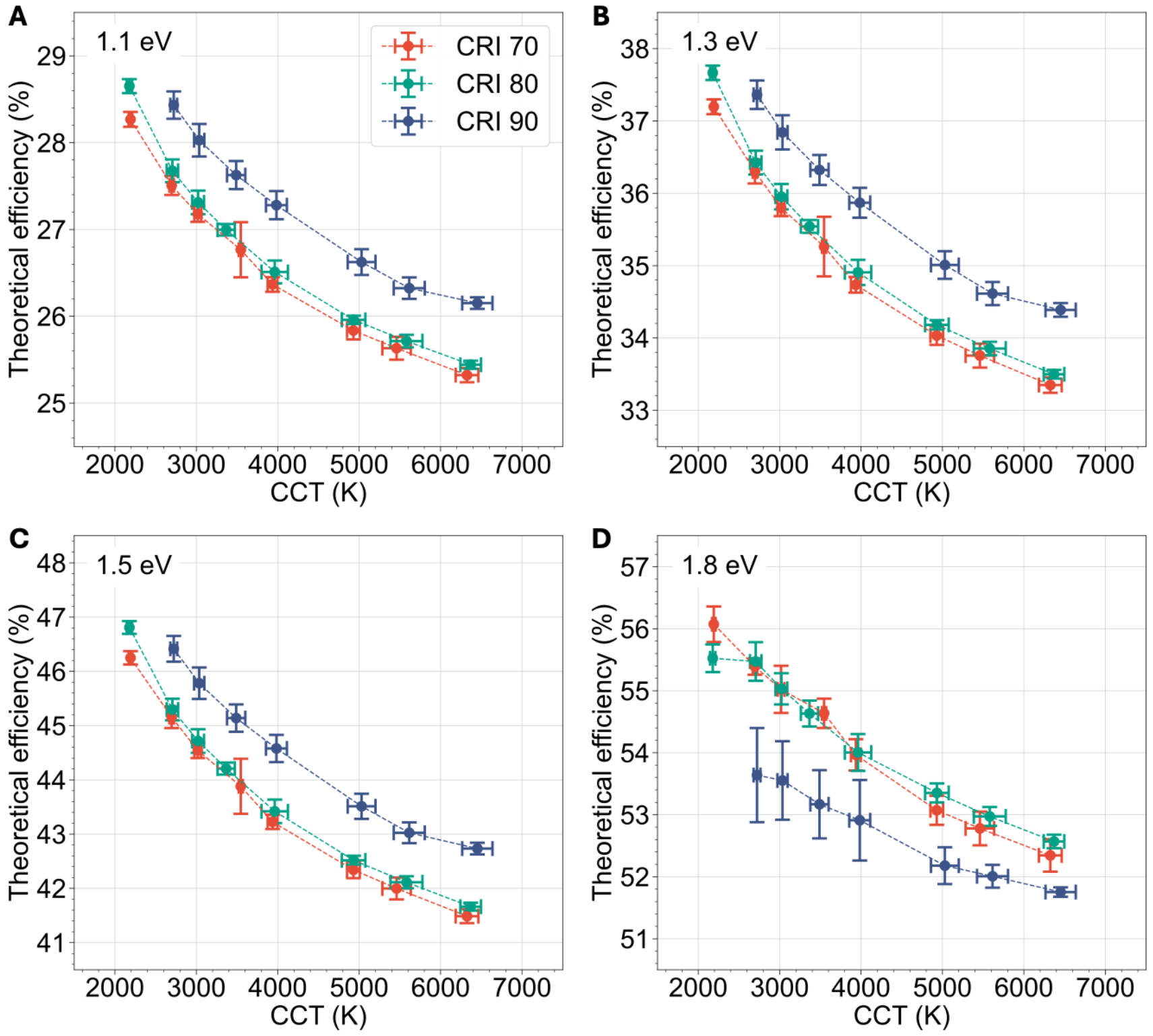

**Figure 4 –** Mean theoretical maximum efficiency and standard deviation as a function of WL-LED color temperature for solar cells with bandgap values of **A,** 1.1 eV, **B,** 1.3 eV, **C,** 1.5 eV, and **D,** 1.8 eV when illuminated by WL-LEDs (Lux = 500 lx) with different CRI values (70, 80, and 90).

As observed in **Fig. 4A-C**, while the magnitude of the theoretical efficiency increases with larger bandgap values as they approach the theoretical optimum, no significant differences in IPV efficiency are, again, observed between using LEDs with CRI of 70 and 80. However, for IPVs with bandgaps between 1.1 to 1.5 eV, the highest theoretical efficiencies are observed for the investigated 90 CRI LEDs due to the extended irradiance spectrum into the red. Therefore, if commercial cells such as c-Si or CdTe are employed as IPVs, while not optimal, IoT devices can, depending on how high CRI is achieved, be powered more efficiently under more natural and vibrant LEDs than under their low CRI counterparts. The opposite trend is observed for IPVs with bandgaps near the optimum (1.8 eV). In this case, the cell is not only more efficient overall but will generally perform better under illumination with the investigated 70 and 80 CRI LEDs. This, once more, highlights the potential application space for wide-bandgap solar cells as IPVs as compared to mature technologies developed for residential and commercial solar applications, and underscores that both CCT and CRI should be accounted for when characterizing IPVs with white-light LEDs as certain high CRI LEDs have broader spectral profiles compared to their lower CRI counterparts.

## CONCLUSION

While prior work has emphasized CCT as the dominant factor affecting IPV bandgap energy and efficiency, we here demonstrate that both CCT and CRI should be accounted for as the broader irradiance spectra of certain high-CRI LEDs shift the optimum bandgap energies to significantly lower values while also slightly reducing the maximum theoretical efficiency. Furthermore, we confirm previous notions that next-generation wide-bandgap materials should offer superior efficiency under typical indoor lighting scenarios, and we highlight that mature technologies (such as c-Si and CdTe) could benefit from illumination with certain high-CRI light. Our work underscores the importance of accounting for both CCT and CRI when developing and characterizing IPVs. These insights can guide IPV deployment in real-world environments such as smart homes, retail, or industrial monitoring, where lighting choices may vary significantly.

## SUPPLEMENTARY MATERIAL

The Supplementary Material contains graphs of all investigated LED spectra, graphs of equivalent calculations shown in the main article, but at fixed irradiance rather than fixed lux, graphs showing how solar characteristics vary as a function of irradiance and illuminance, graphs showing the relationship between irradiance spectrum broadening and CRI, and details of how the actual CCT was calculated from the LED irradiance spectra along with tabulated values for CCT and chromaticity coordinates for each LED spectrum.

## ACKNOWLEDGEMENTS

A.S., I.Y. and J.A.R. would like to thank the Undergraduate Summer Research Program (UGSRP), NYU Tandon School of Engineering for their support. J.A.R. would like to thank Dr. Steven J. Byrnes for writing the foundational code (available at https://sjbyrnes.com/) that was modified for the present study.

## DATA AVAILABILITY STATEMENT

The data that supports the findings of this study are publicly available from online datasheets provided by the LED manufactures but are also within the article [and its supplementary material].

# SUPPLEMENTARY MATERIAL

On the Role of Color Temperature and Color Rendering Index of White-Light LEDs on the Theoretical Efficiency Limit of Indoor Photovoltaics

Aditi Sharma[1], Ian Yum[1], Alexander A. Guaman[1], Gavin Akmam[1], Jason A. Röhr[1,*]

[1]General Engineering, Tandon School of Engineering, New York University, Brooklyn, 11201 New York, United States of America

## Normalized LED spectra

All SPD data were plot digitized manually from publicly available datasheets obtained directly from the suppliers' websites (links provided in captions below). We used the online plotdigitizer tool, plotdigitizer (https://plotdigitizer.com/app). While care was taken, the authors acknowledge that errors could have been introduced during the digitizing process; however, we anticipate these to be minor and not significantly affecting the calculation results presented herein.

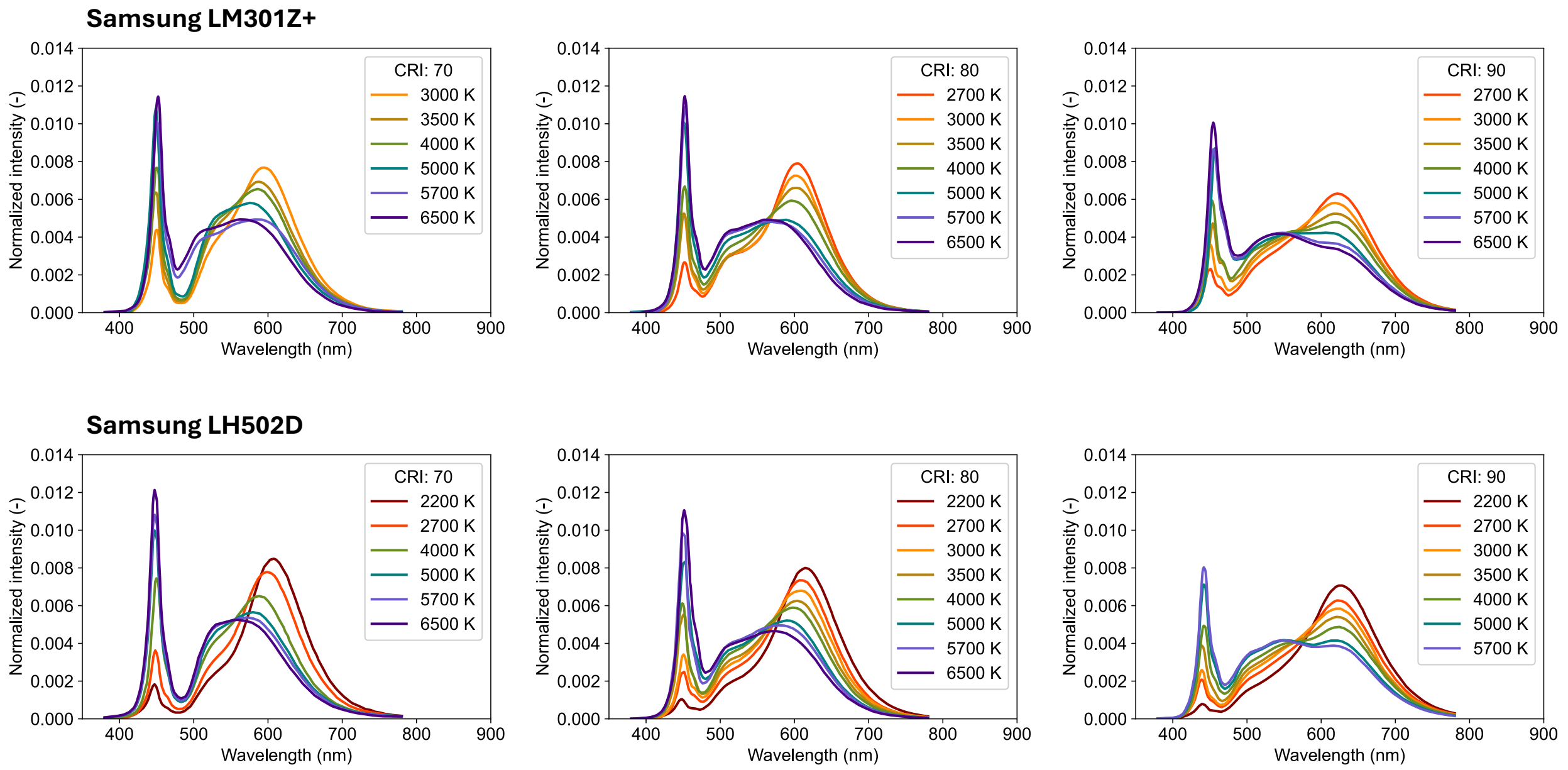

**Figure S5 –** Normalized irradiance spectra of Samsung LM301Z+ (link) and LH502D (link) white-light LEDs.

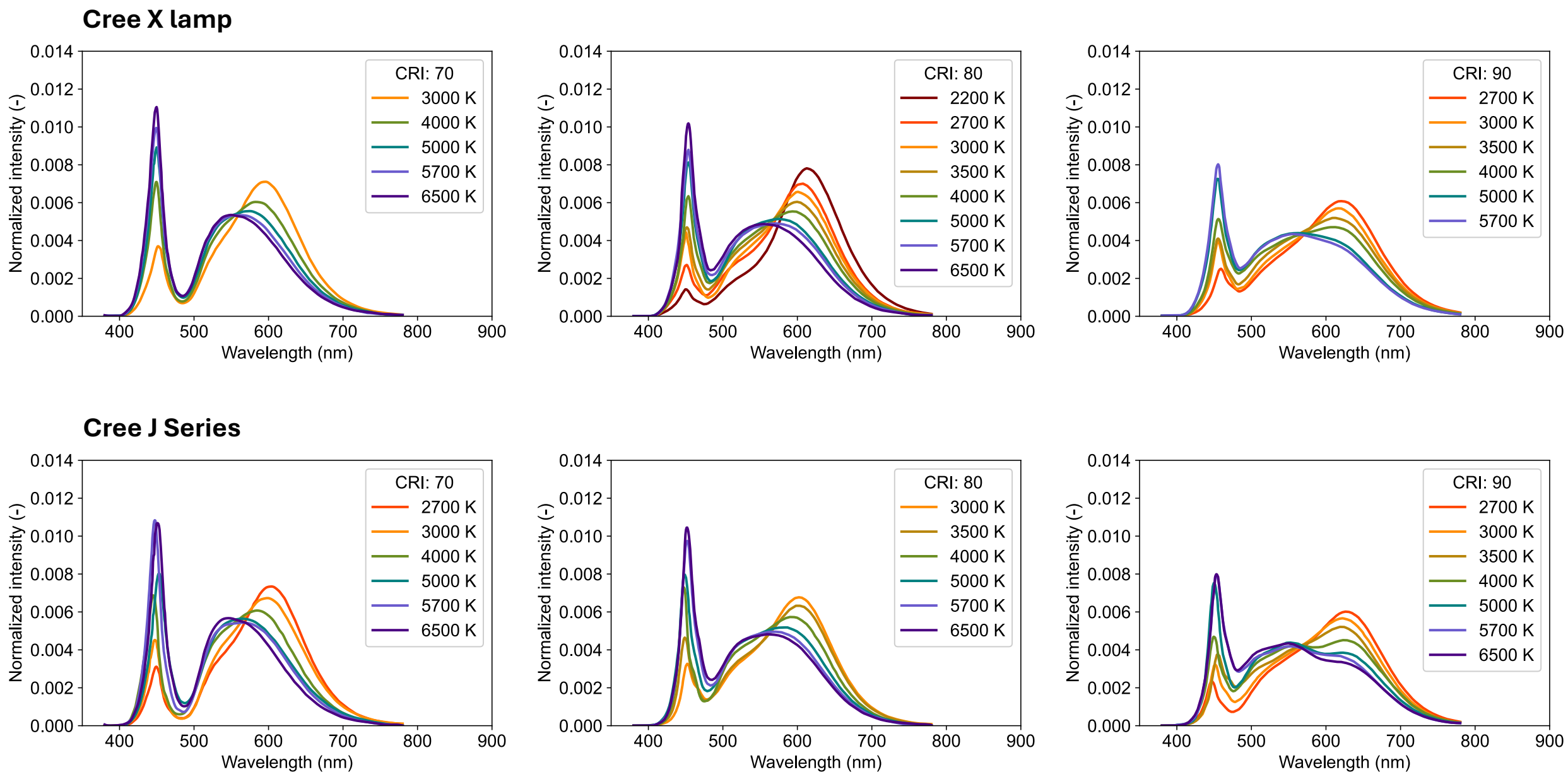

**Figure S6 -** Normalized irradiance spectra of Cree X Lamp (link) and J Series (link) white-light LEDs.

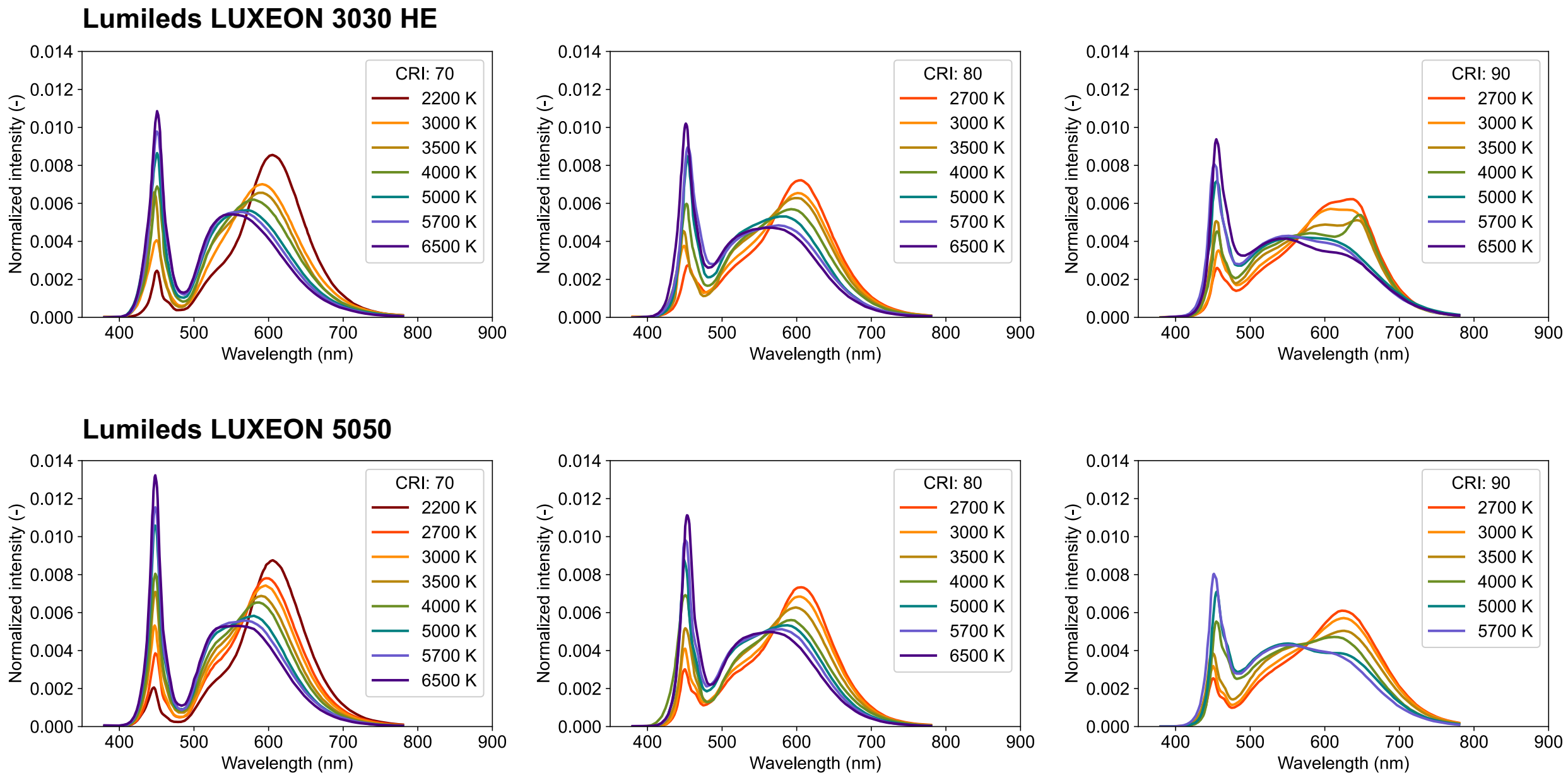

**Figure S3 -** Normalized irradiance spectra of Lumileds LUXEON 3030 HE (link) and 5050 (link) white-light LEDs.

## *Theoretical efficiency and optimum bandgap vs. CCT and CRI*

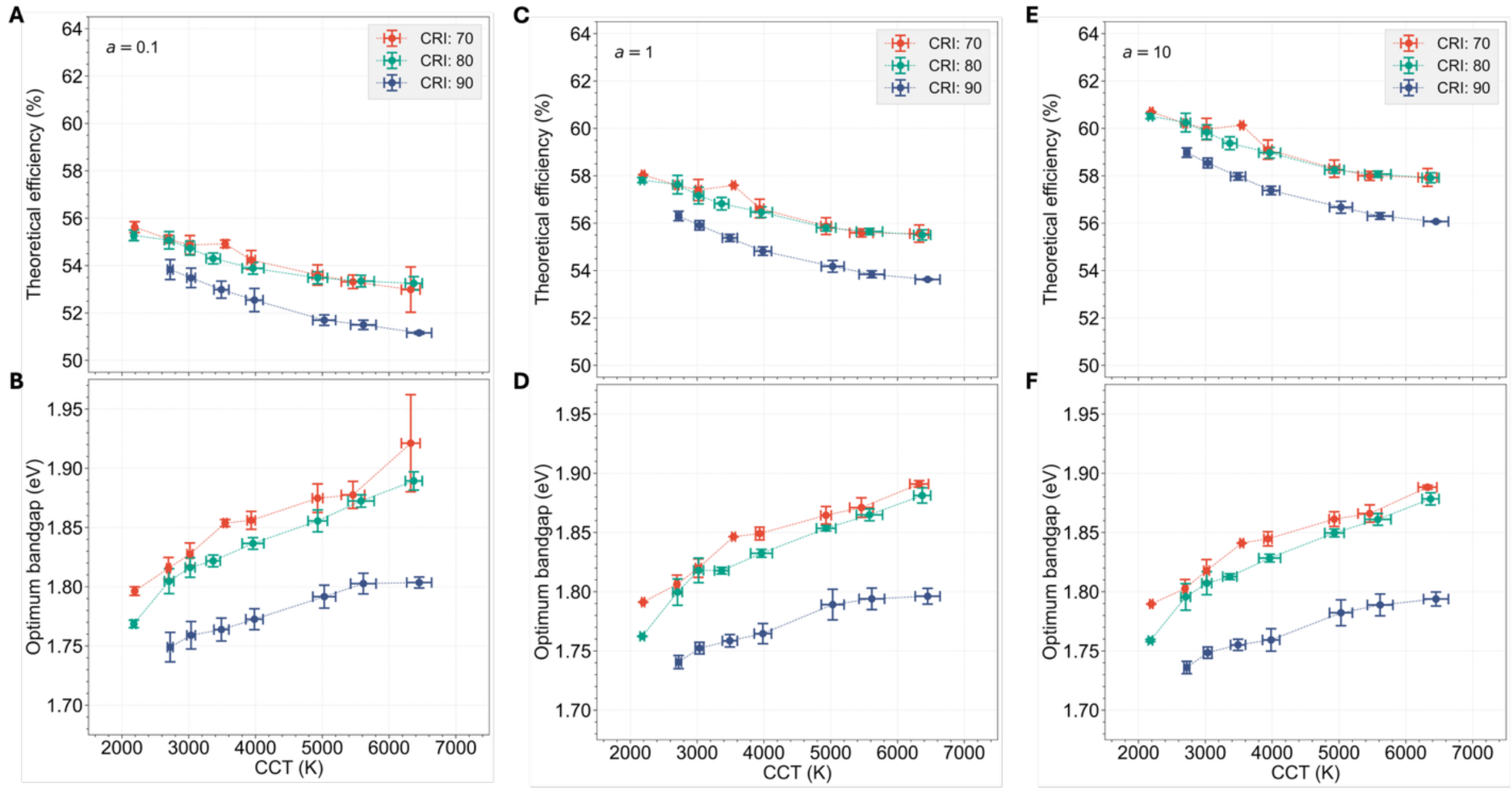

**Figure S4 - A,** Averages and standard deviations of calculated theoretical maximum efficiencies of IPVs illuminated by white-light LEDs (a = 0.1 mW cm$^{-2}$) with different CCT values (ranging from 2200 K to 6500 K) and CRI values (70, 80, and 90). **B,** Corresponding averages of calculated optimum bandgap energies resulting in maximum theoretical efficiency. **C,D,** Similar calculations but for a = 1 mW cm$^{-2}$. **E, F,** Calculations for a = 10 mW cm$^{-2}$.

**$J_{SC}$, $V_{OC}$, FF, and PCE as a function of irradiance or illuminance**

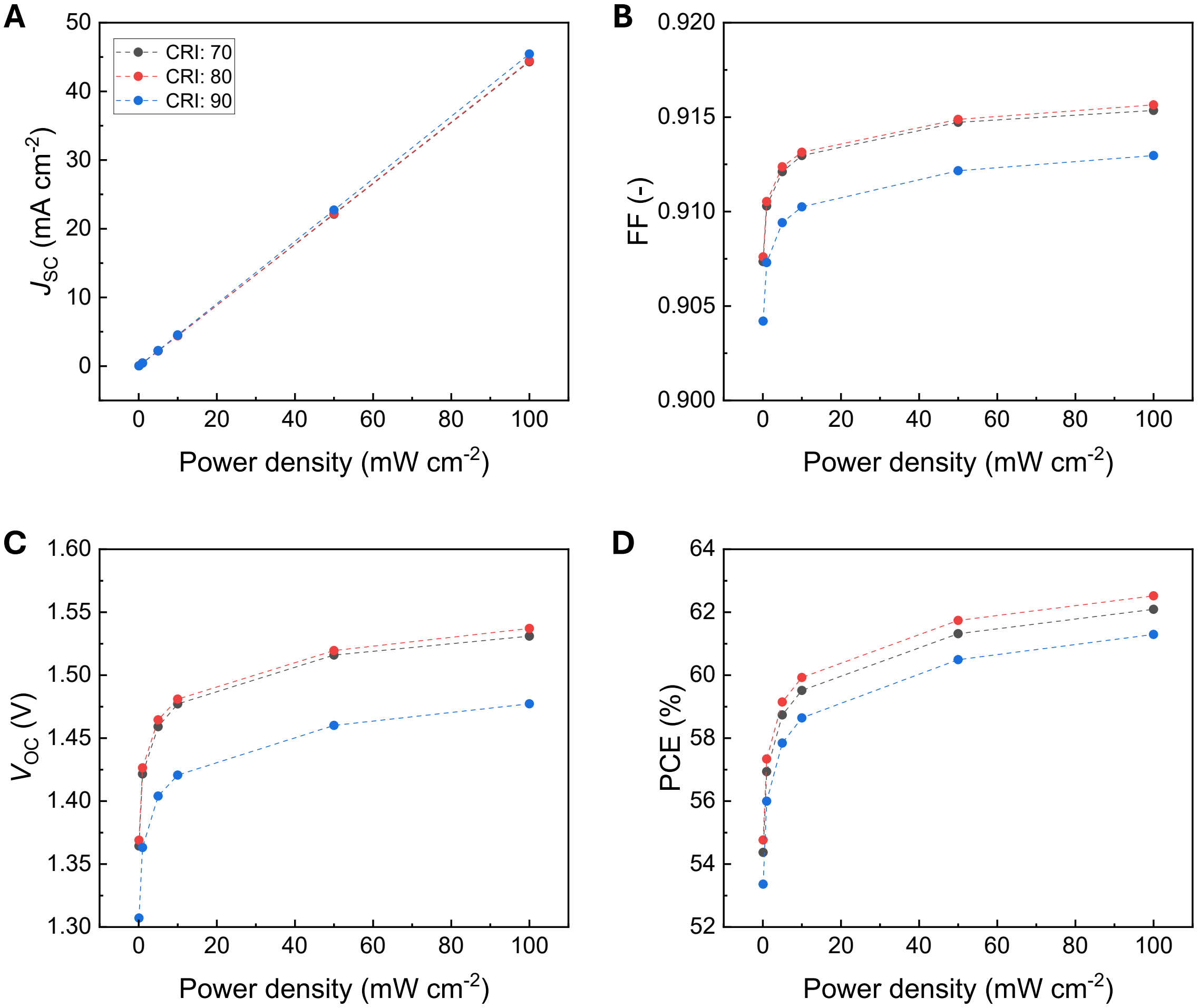

**Figure S5 – A,** $J_{SC}$, **B,** FF, **C,** $V_{OC}$, and **D,** PCE increasing as a function of irradiance power density. The results are shown for calculations performed using the Cree J Series 3000 K LED at 70, 80, and 90 CRI. $E_g$ was set to 1.8 eV for a direct comparison; $T_C$ was set to 300 K. While $J_{SC}$ increases linearly, both FF and $V_{OC}$ increase non-linearly, resulting in a superlinear increase in PCE with increased irradiance.

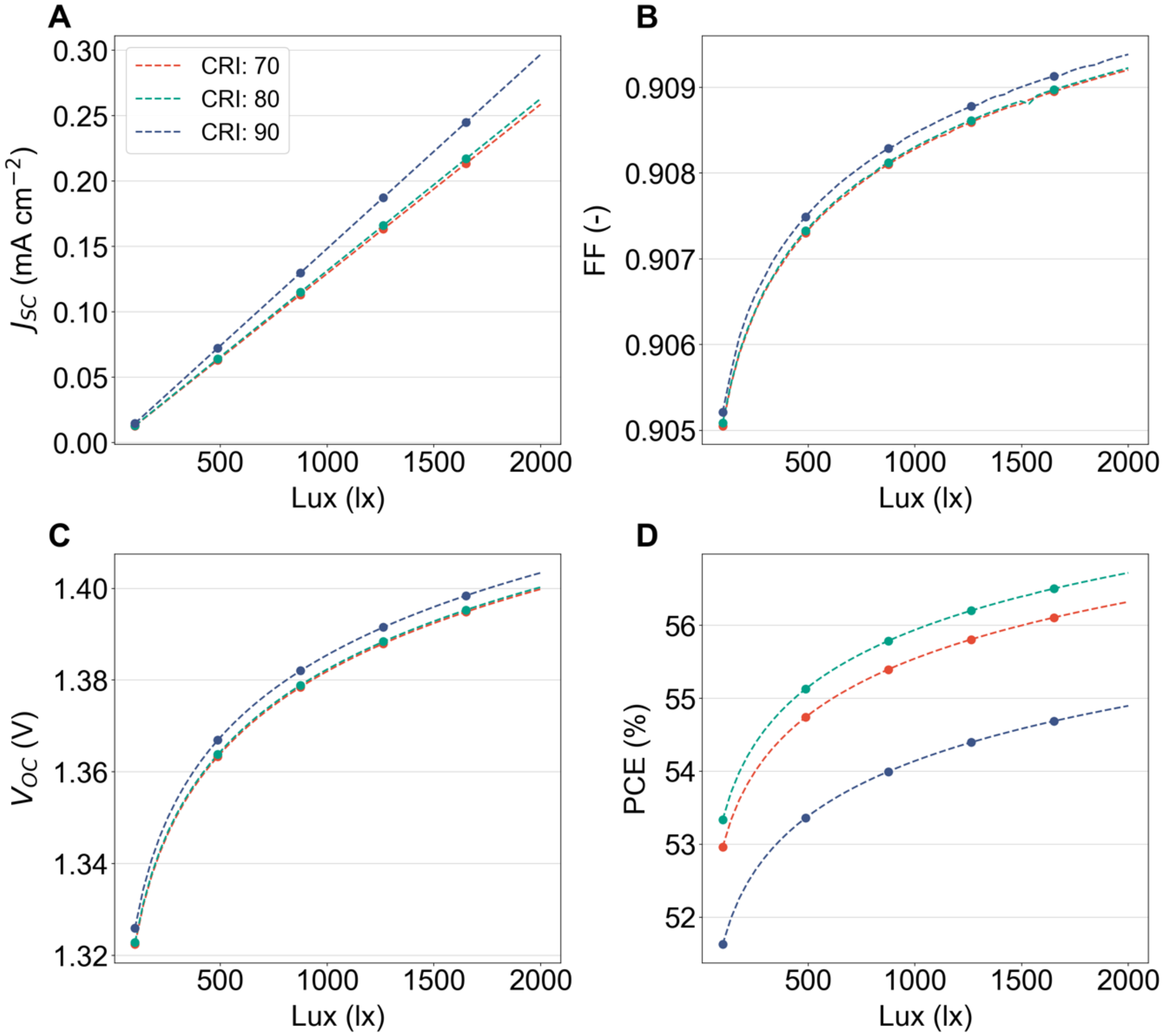

**Figure S6 – A,** $J_{SC}$, **B,** FF, **C,** $V_{OC}$, and **D,** PCE increasing as a function of illuminance. The results are shown for calculations performed using the Cree J Series 3000 K LED at 70, 80, and 90 CRI. $E_g$ was set to 1.8 eV for a direct comparison; $T_C$ was set to 300 K. While $J_{SC}$ increases linearly, both FF and $V_{OC}$ increase non-linearly, resulting in a superlinear increase in PCE with increased illuminance.

## Irradiance spectrum broadening at increased CRI

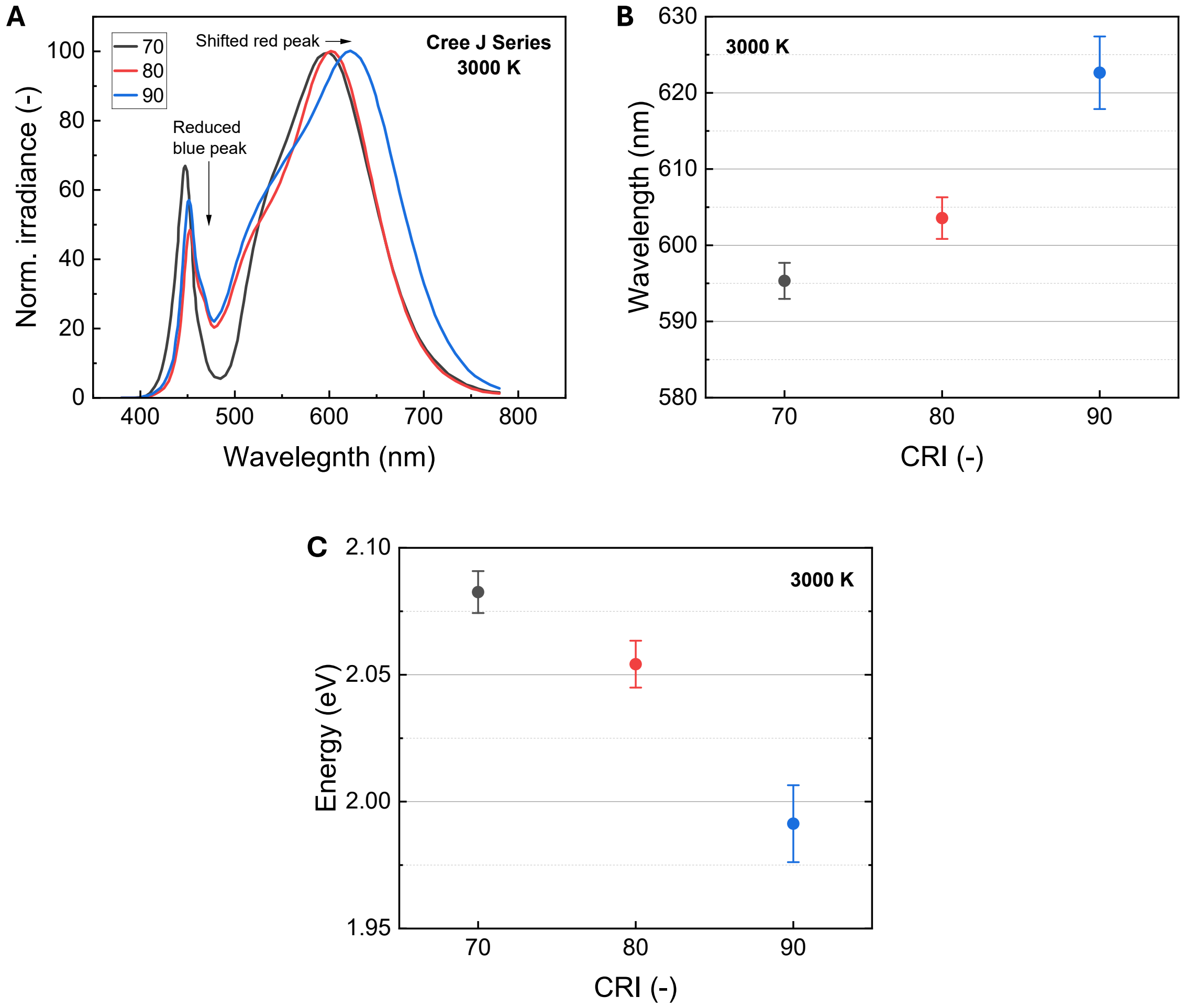

**Figure S7 – A,** Normalized irradiance spectra for Cree J Series LEDs with CCT values of 3000 K and CRI values of 70, 80, and 90, showing how the increased CRI is due to a combination of spectral broadening into the red and increased red to blue irradiance. **B,** Red peak wavelength for all LEDs with CCT of 3000 K. **C,** Red peak energy for all LEDs with CCT of 3000 K.

**Theoretical efficiency vs. CCT and CRI at fixed bandgap values**

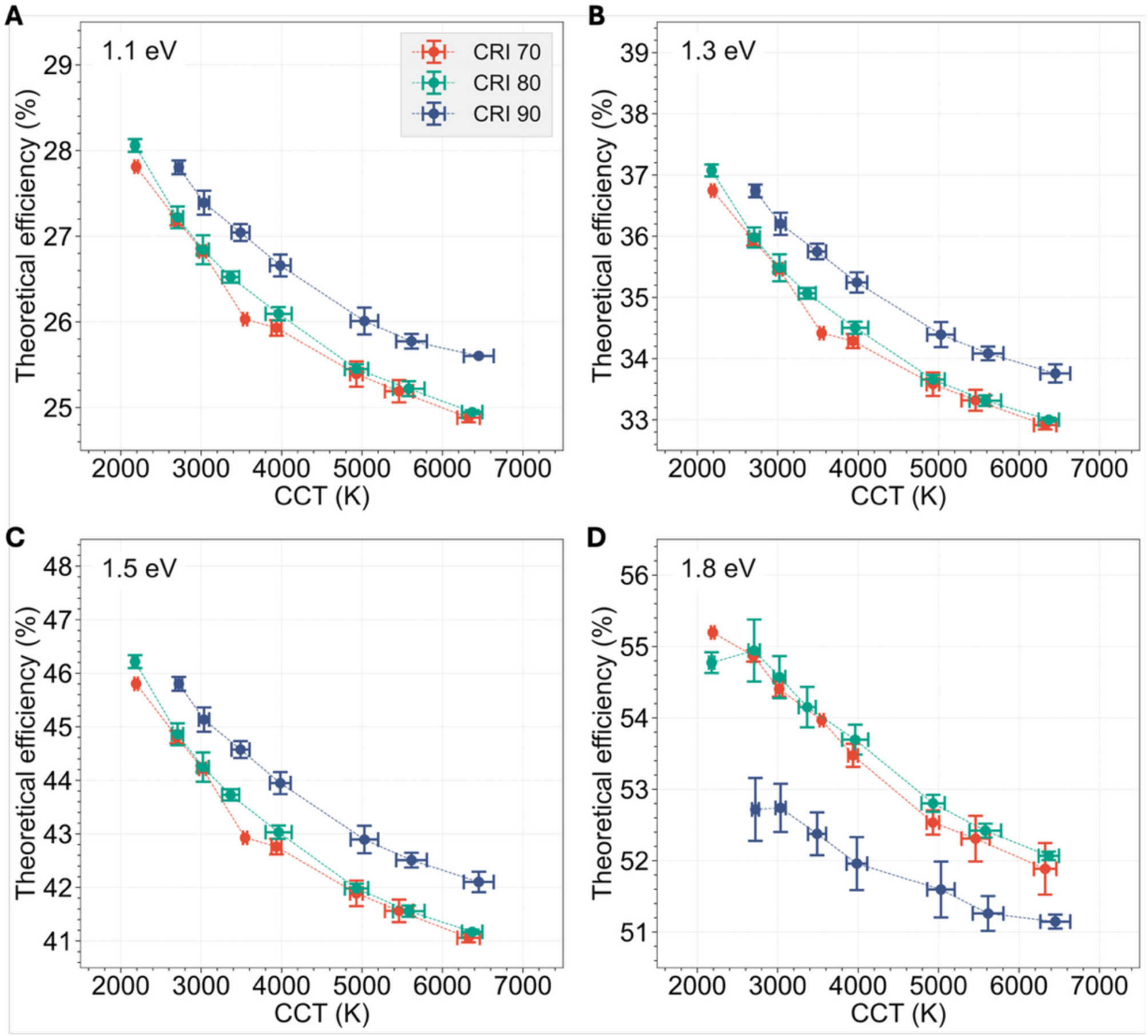

**Figure S8 –** Mean theoretical maximum efficiency and standard deviation as a function of WL-LED color temperature for solar cells with bandgap values of **A,** 1.1 eV, **B,** 1.3 eV, **C,** 1.5 eV, and **D,** 1.8 eV when illuminated by WL-LEDs (a = 0.1 mW cm$^{-2}$) with different CRI values (70, 80, and 90).

**Photopic response curve**

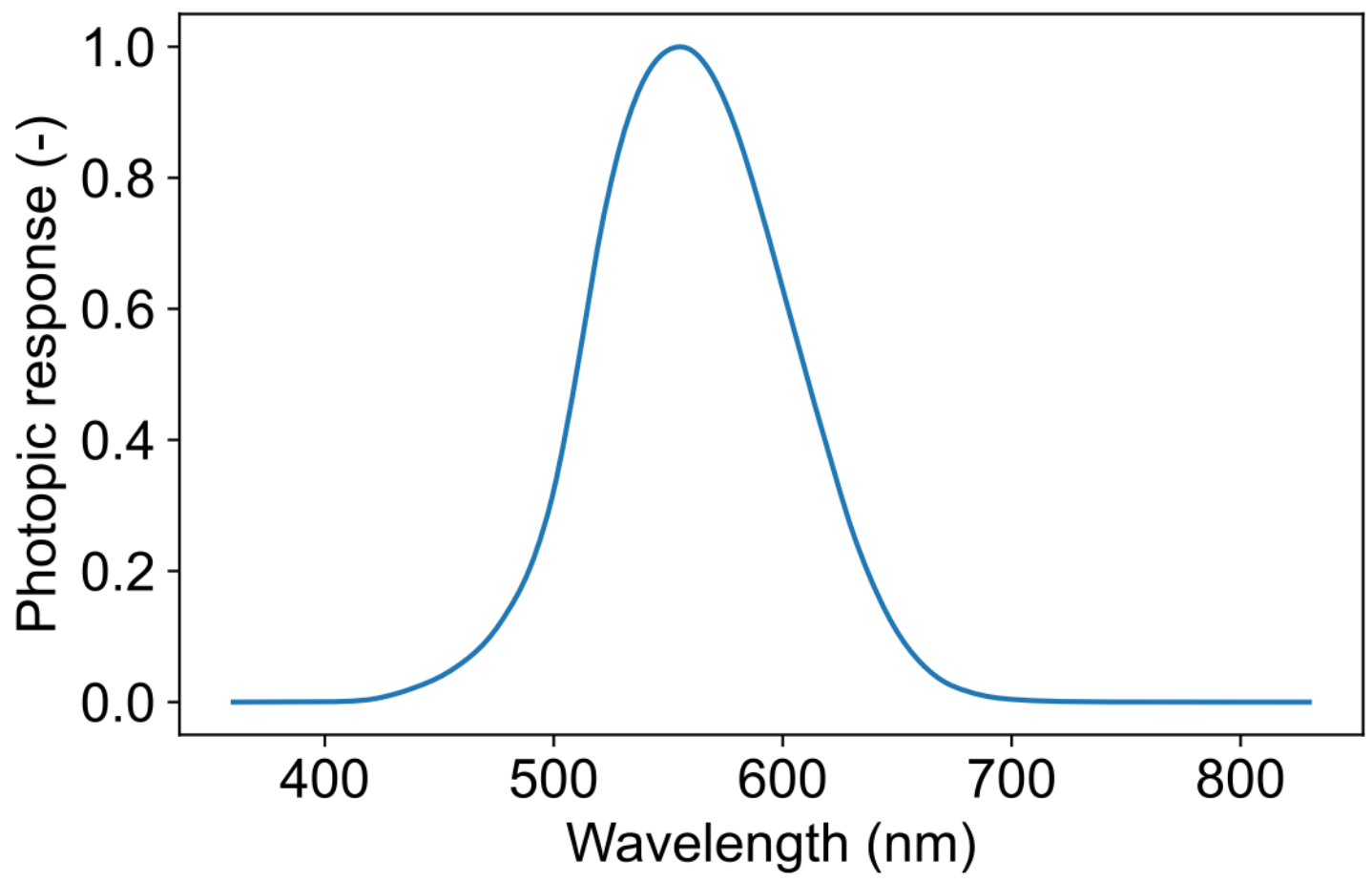

**Figure S9 –** CIE 1931 photopic eye response curve.

**Planckian Black-Body Locus**

The true correlated color temperature (CCT) of a white-light LED irradiance spectrum can differ from nominal color temperature values ($CCT_0$) listed in the manufacturer data sheets. To calculate the true CCT, we need to compare the normalized irradiance spectra (**Fig. S1-S3**) to the Planckian black-body (BB) locus, which is the color that a black body would have in chromaticity space as the temperature changes. To calculate the BB locus, we first define Planck's spectral irradiance as,

$$\varphi_{\mathrm{BB}}(\lambda, T) = \frac{2hc^2}{\lambda^5} \frac{1}{\exp\left(\frac{hc}{\lambda k_{\mathrm{B}} T}\right) - 1} \tag{S1}$$

where $h$ is Planck's constant, $c$ is the speed of light, $\lambda$ is the frequency of light, $k_{\mathrm{B}}$ is Boltzmann's constant, and $T$ is the temperature of the black body. From this, we can calculate the *International Commission on Illumination* (CIE) tristimulus values for the BB locus from,

$$X_{\mathrm{BB}}(T) = \int_{\lambda_{\min}}^{\lambda_{\max}} \varphi_{\mathrm{BB}}(\lambda, T) \times \bar{x}\, d\lambda \tag{S2}$$

$$Y_{\mathrm{BB}}(T) = \int_{\lambda_{\min}}^{\lambda_{\max}} \varphi_{\mathrm{BB}}(\lambda, T) \times \bar{y}\, d\lambda \tag{S3}$$

$$Z_{\mathrm{BB}}(T) = \int_{\lambda_{\min}}^{\lambda_{\max}} \varphi_{\mathrm{BB}}(\lambda, T) \times \bar{z}\, d\lambda \tag{S4}$$

where $\bar{x}$, $y$, and $\bar{z}$ are the CIE 1931 2° color matching functions (shown in **Fig. S10**). These functions were obtained from the Python package, *Colour*, which will also used for its CCT calculation algorithms and for graphing the chromaticity charts. The documentation for *Colour* can be found on Zenodo: https://doi.org/10.5281/zenodo.13917514.

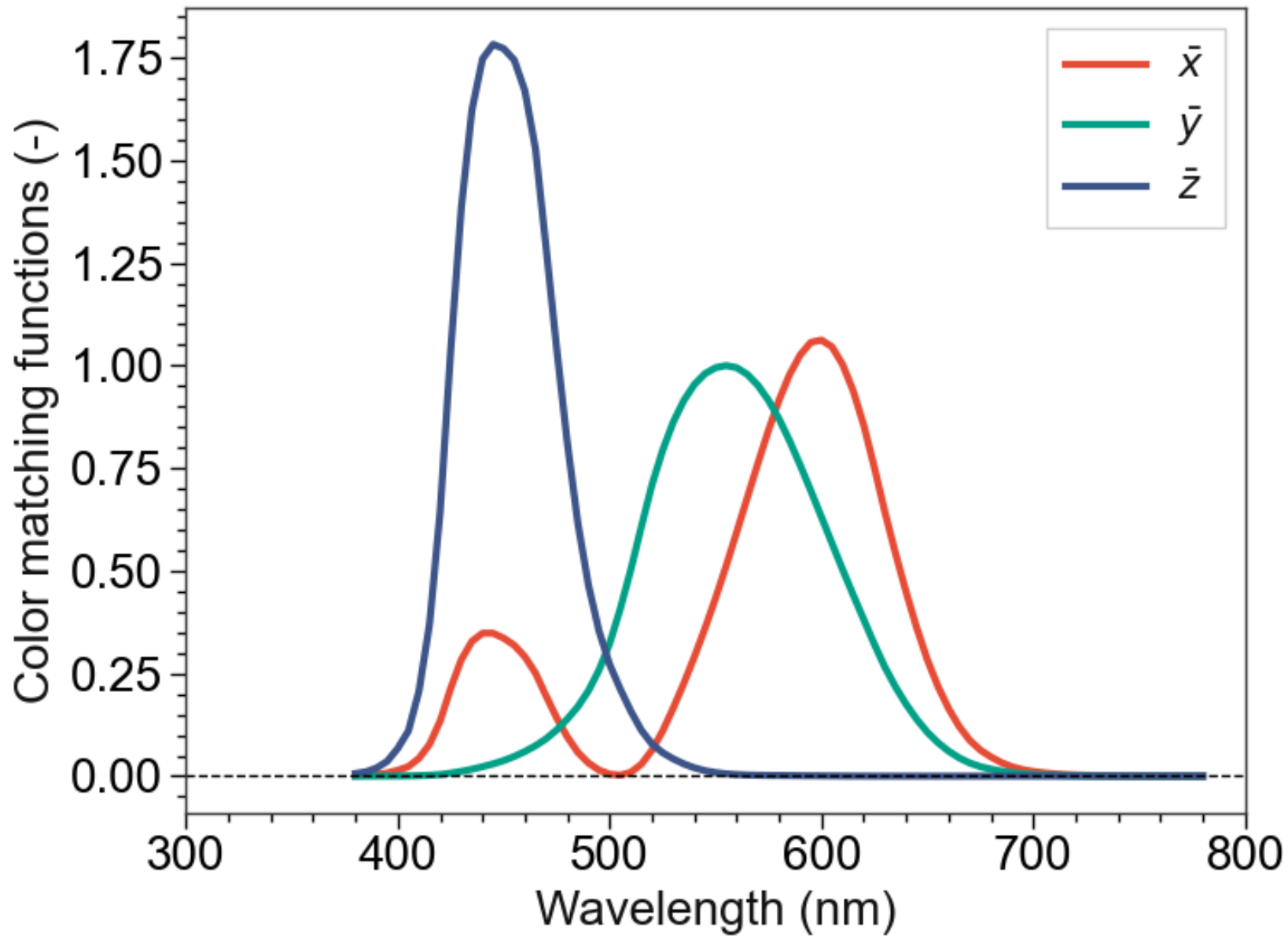

**Figure S10 –** The CIE 1931 2° color matching functions.

From the tristimulus values (**Eq. S2-S4**) we can calculate the CIE 1931 chromaticity coordinates,

$$x_{\mathrm{BB}}(T) = \frac{X_{\mathrm{BB}}(T)}{X_{\mathrm{BB}}(T) + Y_{\mathrm{BB}}(T) + Z_{\mathrm{BB}}(T)} \tag{S5}$$

$$y_{\mathrm{BB}}(T) = \frac{Y_{\mathrm{BB}}(T)}{X_{\mathrm{BB}}(T) + Y_{\mathrm{BB}}(T) + Z_{\mathrm{BB}}(T)} \tag{S6}$$

which allows us to project the BB locus onto the CIE 1931 chromaticity chart, as shown in **Fig. S11**.

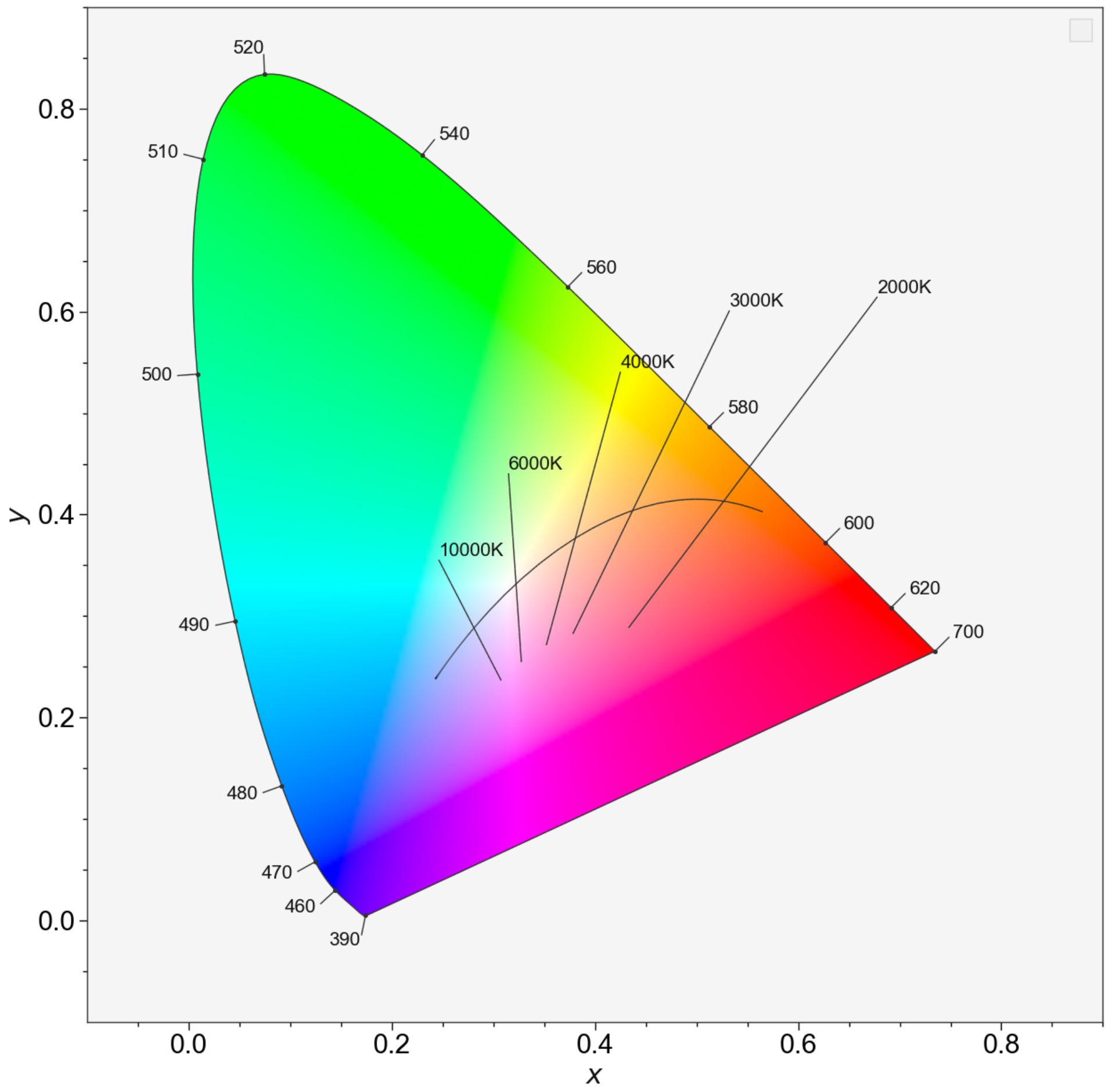

**Figure S11 –** The Planckian black-body locus (from 1667 K to 25000 K) overlaid on the CIE 1931 chromaticity chart. CT ticks for 2000 K, 3000 K, 4000 K, 6000 K, and 10000 K are shown.

**LED chromaticity coordinates**

To compare the BB locus to the LED chromaticity coordinates, which will also be used to calculate the CCT for each LED irradiance spectrum (**Fig. S1-S3**), we use the normalized irradiance spectra to calculate the tristimulus values,

$$X = \int_{\lambda_{\min}}^{\lambda_{\max}} \varphi_{\text{norm}}(\lambda) \times \bar{x}(\lambda)\, d\lambda \tag{S7}$$

$$Y = \int_{\lambda_{\min}}^{\lambda_{\max}} \varphi_{\text{norm}}(\lambda) \times \bar{y}(\lambda)\, d\lambda \tag{S8}$$

$$Z = \int_{\lambda_{\min}}^{\lambda_{\max}} \varphi_{\text{norm}}(\lambda) \times \bar{z}(\lambda)\, d\lambda \tag{S9}$$

From these tristimulus values, we can calculate the corresponding chromaticity coordinates,

$$x = \frac{X}{X + Y + Z} \tag{S10}$$

$$y = \frac{Y}{X + Y + Z} \tag{S11}$$

which allows us to project the LED chromaticity onto the CIE 1931 chromaticity chart in comparison with the BB locus. We show charts for each set of LEDs below (**Fig. S12-S29**). Chromaticity coordinates for all LEDs are listed in **Table S1-18**.

**CCT Calculations**

To calculate the CCT values, we convert the $X, Y, Z$ coordinates into CIE 1960 UCS $u, v$ (along with the corresponding temperature-dependent $u_{\mathrm{BB}}, v_{\mathrm{BB}}$) chromaticity coordinates,

$$u = \frac{4X}{X + 15Y + 3Z} \quad \text{and} \quad v = \frac{6Y}{X + 15Y + 3Z} \tag{S12}$$

$$u_{\mathrm{BB}} = \frac{4X_{\mathrm{BB}}}{X_{\mathrm{BB}} + 15Y_{\mathrm{BB}} + 3Z_{\mathrm{BB}}} \quad \text{and} \quad v_{\mathrm{BB}} = \frac{6Y_{\mathrm{BB}}}{X_{\mathrm{BB}} + 15Y_{\mathrm{BB}} + 3Z_{\mathrm{BB}}} \tag{S13}$$

To find CCT, *Colour* uses the following expression,

$$D = \sqrt{[u - u_{\mathrm{BB}}(T)]^2 + [v - v_{\mathrm{BB}}(T)]^2} \tag{S14}$$

where the problem is to find the value for $T$ which results in the lowest value for $D$, which in turn is then the CCT. CCT values and chromaticity coordinates for all LEDs are listed in **Table S1-18**.

**$\Delta uv$ Calculations**

Many white-light sources, such as LEDs, do not sit perfectly on the BB locus which can result in the light (tint) appearing greener or more pink in color even if they have the same CCT. To tabulate this, once $u_{\mathrm{BB}}(T)$ and $v_{\mathrm{BB}}(T)$ have been identified (by finding the corresponding CCT), we use *Colour* to calculate the distance from the locus (i.e., the lines extending from the BB locus in **Fig. S11**), $\Delta uv$, using,

$$\Delta uv = \frac{\frac{du}{dT}[v - v_{\mathrm{BB}}(T)] - \frac{dv}{dT}[u - u_{\mathrm{BB}}(T)]}{\sqrt{\left[\frac{du}{dT}\right]^2 + \left[\frac{dv}{dT}\right]^2}} \tag{S15}$$

Positive $\Delta uv$ means that the color appears greener while a negative $\Delta uv$ means that the color appears pinker. While human vision is highly sensitive, for $|\Delta uv| \leq 0.001$ the color difference is considered negligible, and the light source is essentially indistinguishable from a true BB (white-light) emitter.

## Color-temperature characteristics of the LED spectra

Below, $CCT_0$, CCT, ΔCCT = CCT – $CCT_0$, $x$, $y$, $u$, $v$, and $\Delta uv$ results are tabulated (**Table S1-18**) for each LED spectrum investigated within this study. Additionally, we show overlays of the LED chromaticity coordinates on CIE 1931 chromaticity charts in comparison with the BB locus (**Fig. S12-29**).

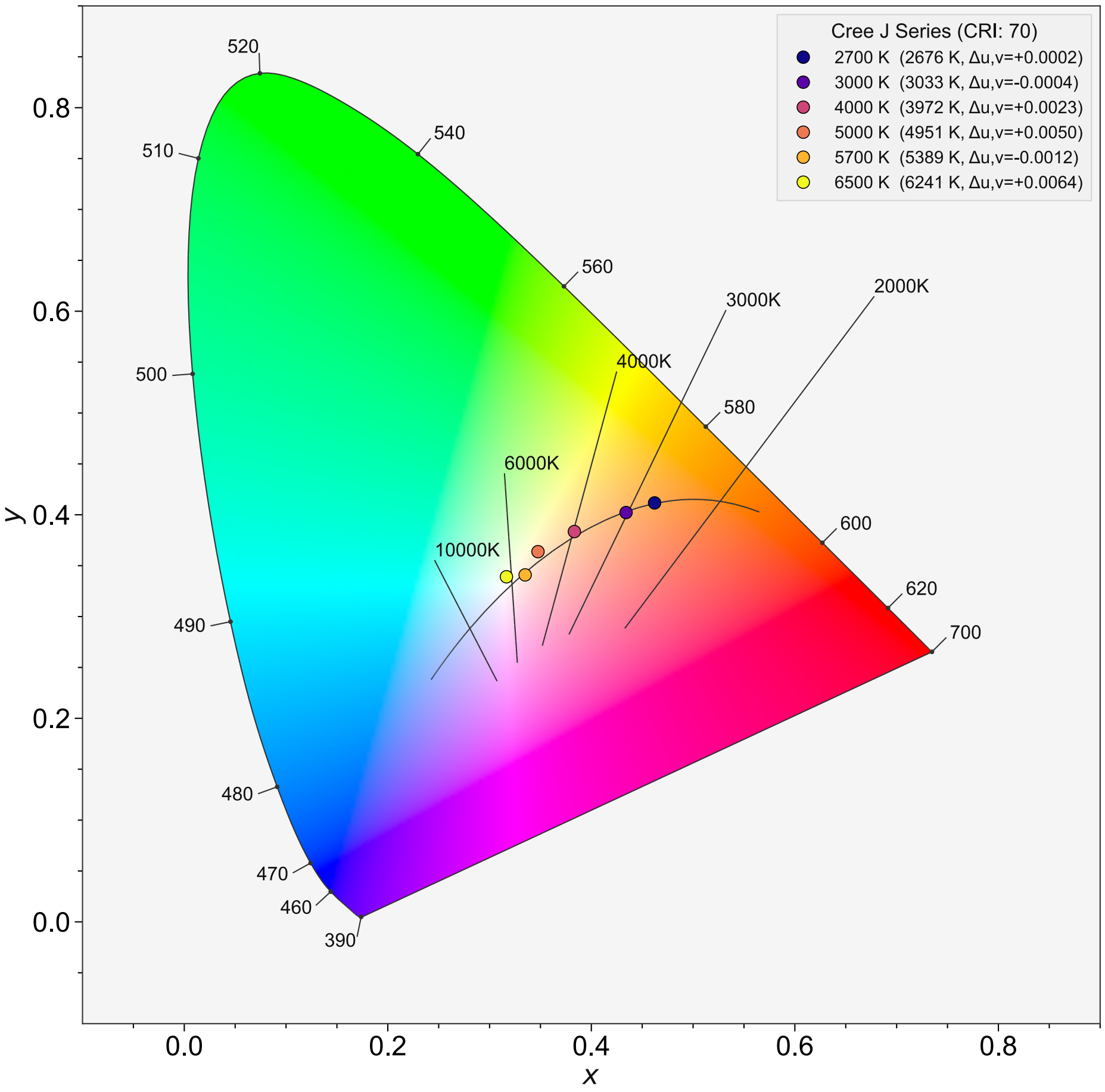

**Figure S12 –** Chromaticity chart for the 70 CRI Cree J Series LEDs.

**Table S1 –** Correlated color temperature values and chromaticity coordinates for the 70 CRI Cree J Series LEDs.

| **Cree J Series (CRI: 70)** | | | | | | | |
|---|---|---|---|---|---|---|---|
| $CCT_0$ | CCT | ΔCCT | $x$ | $y$ | $u$ | $v$ | $\Delta uv$ |
| 2200 K | - | - | - | - | - | - | - |
| 2700 K | 2676.30 K | -23.70 K | 0.4622 | 0.4117 | 0.2635 | 0.3521 | 0.000201 |
| 3000 K | 3032.61 K | 32.61 K | 0.4341 | 0.4022 | 0.2496 | 0.3468 | -0.000360 |
| 3500 K | - | - | - | - | - | - | - |
| 4000 K | 3971.95 K | -28.05 K | 0.3833 | 0.3835 | 0.2243 | 0.3366 | 0.002298 |
| 5000 K | 4950.97 K | -49.03 K | 0.3476 | 0.3638 | 0.2084 | 0.3272 | 0.005029 |
| 5700 K | 5389.50 K | -310.50 K | 0.3350 | 0.3409 | 0.2087 | 0.3186 | -0.001171 |
| 6500 K | 6240.95 K | -259.05 K | 0.3166 | 0.3391 | 0.1968 | 0.3161 | 0.006389 |

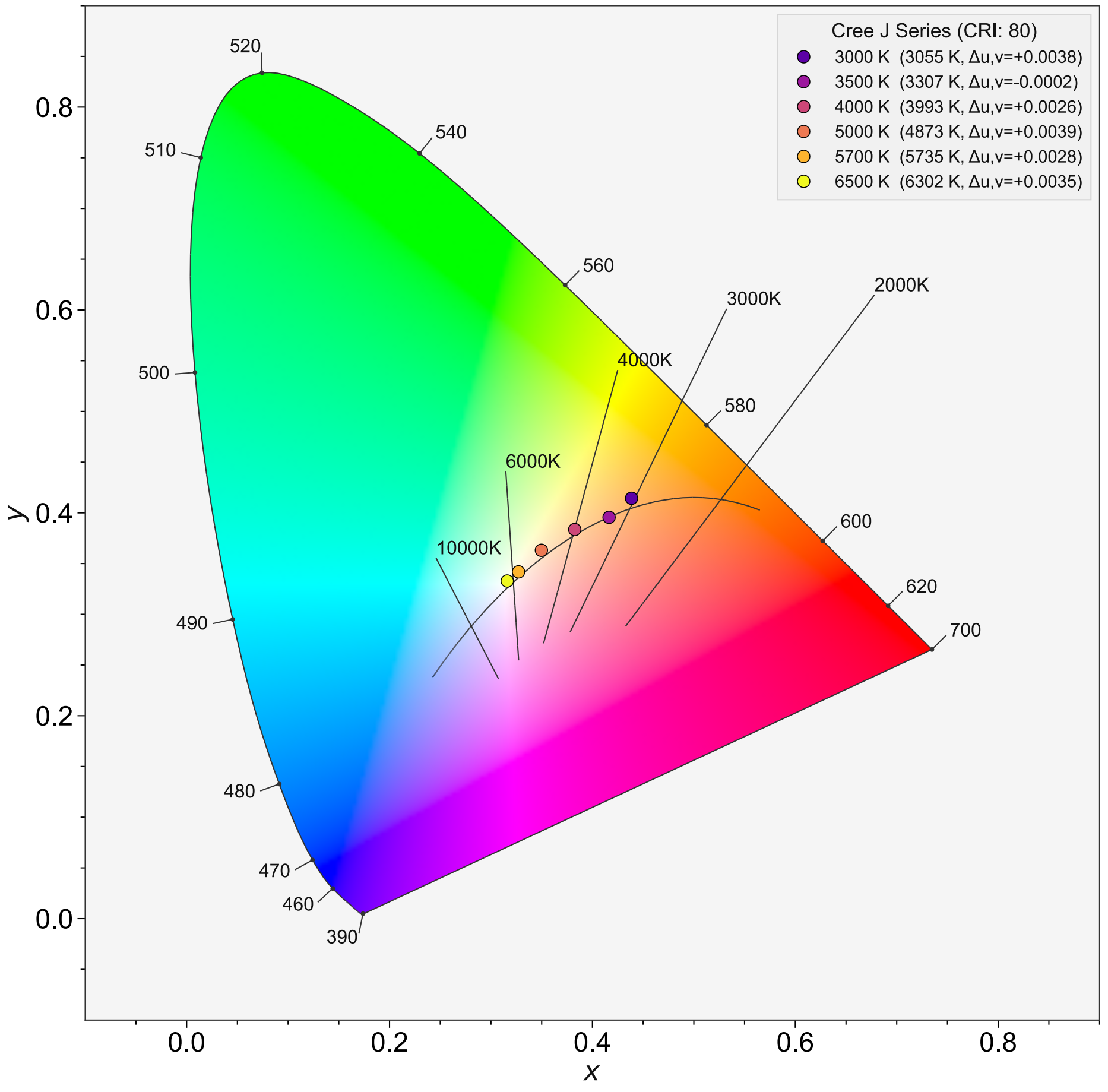

**Figure S13 –** Chromaticity chart for the 80 CRI Cree J Series LEDs.

**Table S2 -** Correlated color temperature values and chromaticity coordinates for the 80 CRI Cree J Series LEDs.

| **Cree J Series (CRI: 80)** | | | | | | | |
|---|---|---|---|---|---|---|---|
| $CCT_0$ | CCT | ΔCCT | $x$ | $y$ | $u$ | $v$ | $\Delta uv$ |
| 2200 K | - | - | - | - | - | - | - |
| 2700 K | - | - | - | - | - | - | - |
| 3000 K | 3054.96 K | 54.96 K | 0.4386 | 0.4144 | 0.2472 | 0.3504 | 0.003832 |
| 3500 K | 3307.16 K | -192.84 K | 0.4164 | 0.3956 | 0.2409 | 0.3433 | -0.000165 |
| 4000 K | 3993.36 K | -6.64 K | 0.3825 | 0.3836 | 0.2238 | 0.3366 | 0.002557 |
| 5000 K | 4873.29 K | -126.71 K | 0.3497 | 0.3631 | 0.2101 | 0.3272 | 0.003874 |
| 5700 K | 5734.83 K | 34.83 K | 0.3272 | 0.3418 | 0.2030 | 0.3181 | 0.002793 |
| 6500 K | 6301.78 K | -198.22 K | 0.3161 | 0.3328 | 0.1987 | 0.3139 | 0.003490 |

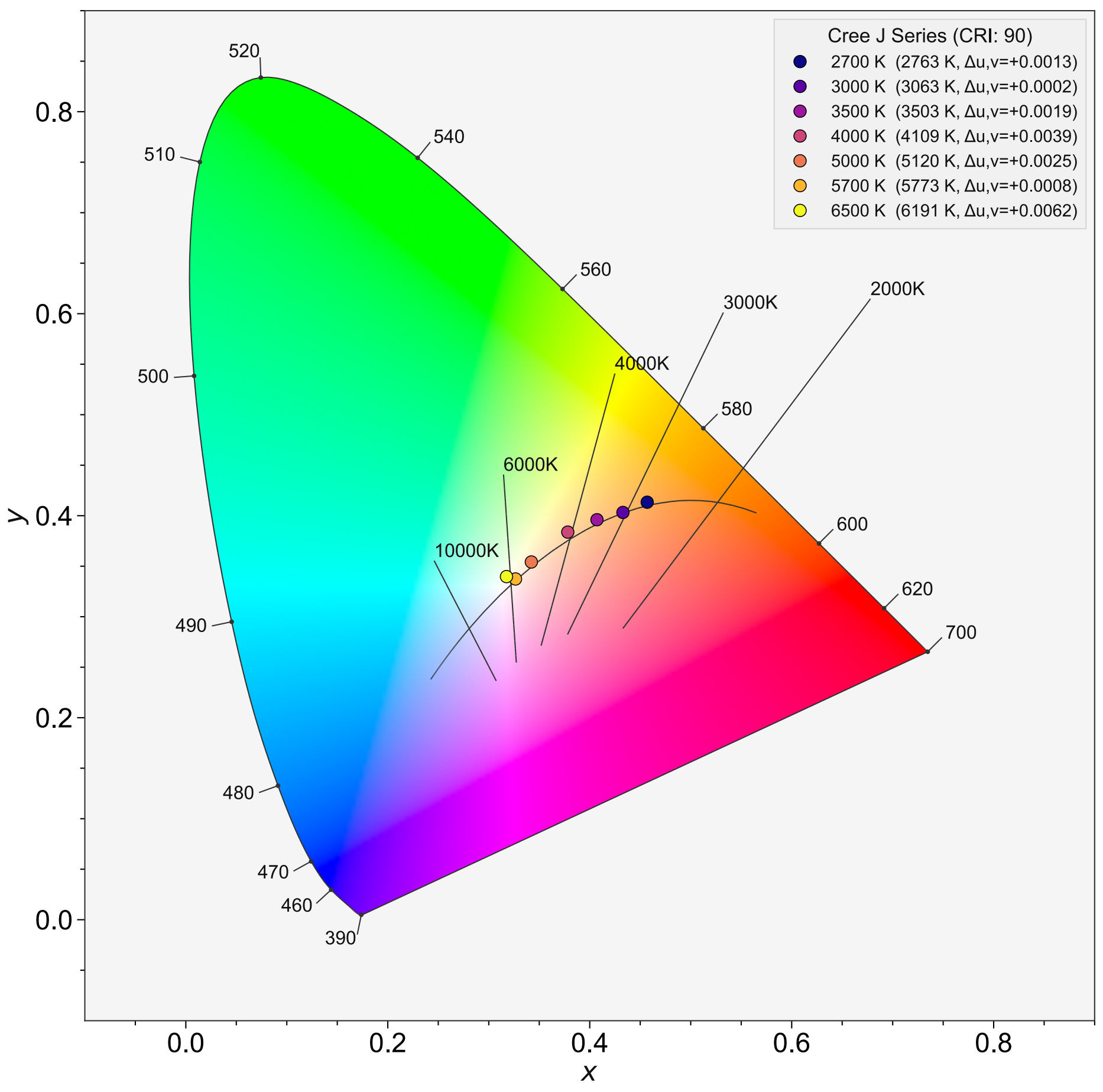

**Figure S14 –** Chromaticity chart for the 90 CRI Cree J Series LEDs.

**Table S3 -** Correlated color temperature values and chromaticity coordinates for the 90 CRI Cree J Series LEDs.

| **Cree J Series (CRI: 90)** | | | | | | | |
|---|---|---|---|---|---|---|---|
| CCT$_0$ | CCT | ΔCCT | $x$ | $y$ | $u$ | $v$ | $\Delta uv$ |
| 2200 K | - | - | - | - | - | - | - |
| 2700 K | 2763.38 K | 63.38 K | 0.4569 | 0.4134 | 0.2594 | 0.3520 | 0.001282 |
| 3000 K | 3062.57 K | 62.57 K | 0.4329 | 0.4032 | 0.2483 | 0.3470 | 0.000237 |
| 3500 K | 3502.60 K | 2.60 K | 0.4072 | 0.3960 | 0.2347 | 0.3425 | 0.001906 |
| 4000 K | 4109.34 K | 109.34 K | 0.3782 | 0.3838 | 0.2209 | 0.3362 | 0.003904 |
| 5000 K | 5120.20 K | 120.20 K | 0.3422 | 0.3542 | 0.2085 | 0.3237 | 0.002476 |
| 5700 K | 5772.87 K | 72.87 K | 0.3265 | 0.3374 | 0.2042 | 0.3165 | 0.000819 |
| 6500 K | 6190.63 K | -309.37 K | 0.3176 | 0.3397 | 0.1972 | 0.3164 | 0.006232 |

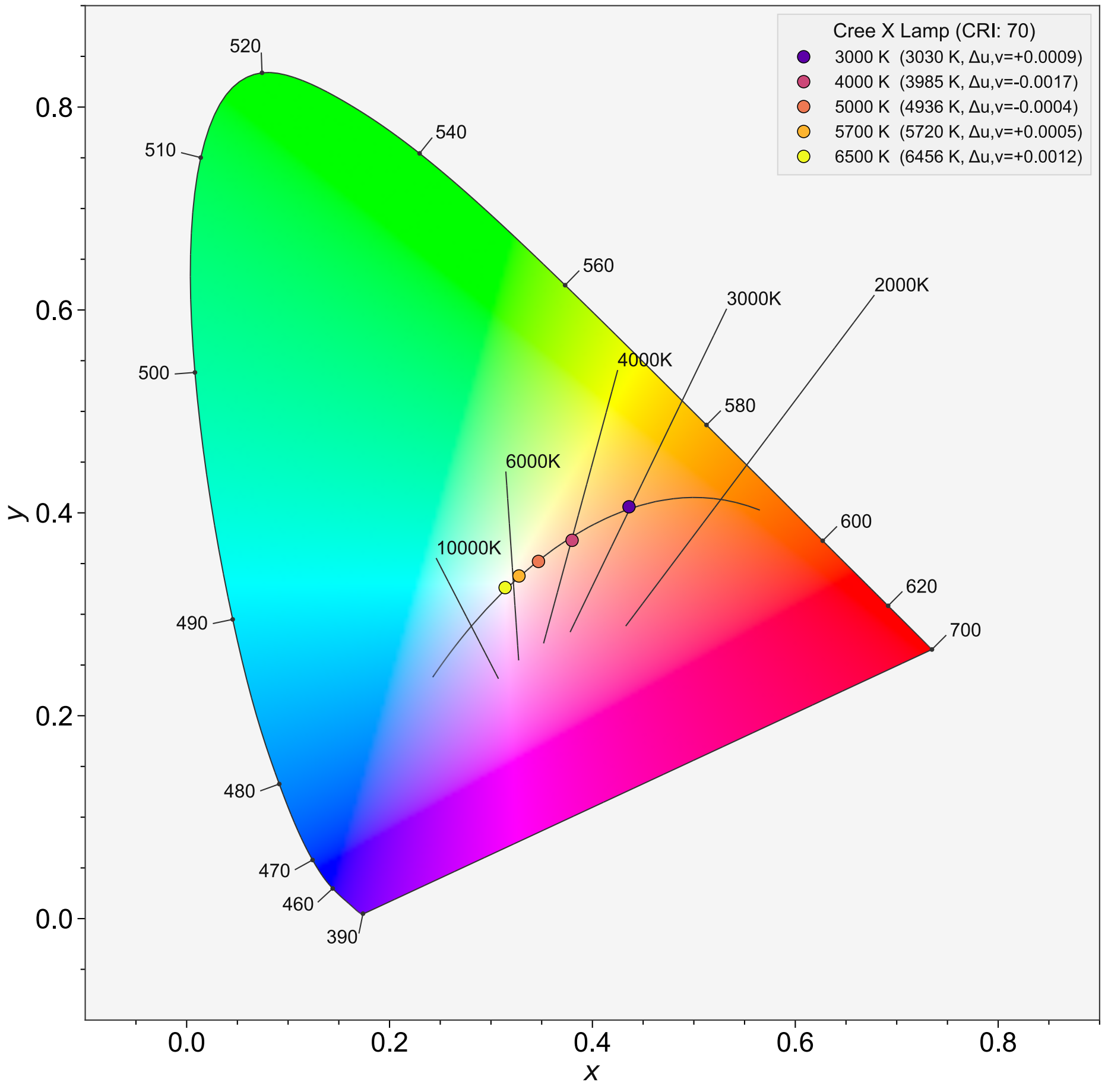

**Figure S15 –** Chromaticity chart for the 70 CRI Cree X Lamp LEDs.

**Table S4 -** Correlated color temperature values and chromaticity coordinates for the 70 CRI Cree XLamp LEDs.

| **Cree Xlamp (CRI: 70)** | | | | | | | |
|---|---|---|---|---|---|---|---|
| $CCT_0$ | CCT | ΔCCT | $x$ | $y$ | $u$ | $v$ | $\Delta uv$ |
| 2200 K | - | - | - | - | - | - | - |
| 2700 K | - | - | - | - | - | - | - |
| 3000 K | 3030.11 K | 30.11 K | 0.4361 | 0.4060 | 0.2492 | 0.3480 | 0.000880 |
| 3500 K | - | - | - | - | - | - | - |
| 4000 K | 3985.09 K | -14.91 K | 0.3799 | 0.3729 | 0.2263 | 0.3332 | -0.001654 |
| 5000 K | 4936.07 K | -63.93 K | 0.3468 | 0.3521 | 0.2124 | 0.3235 | -0.000436 |
| 5700 K | 5719.60 K | 19.60 K | 0.3276 | 0.3378 | 0.2048 | 0.3168 | 0.000511 |
| 6500 K | 6455.93 K | -44.07 K | 0.3140 | 0.3263 | 0.1997 | 0.3114 | 0.001177 |

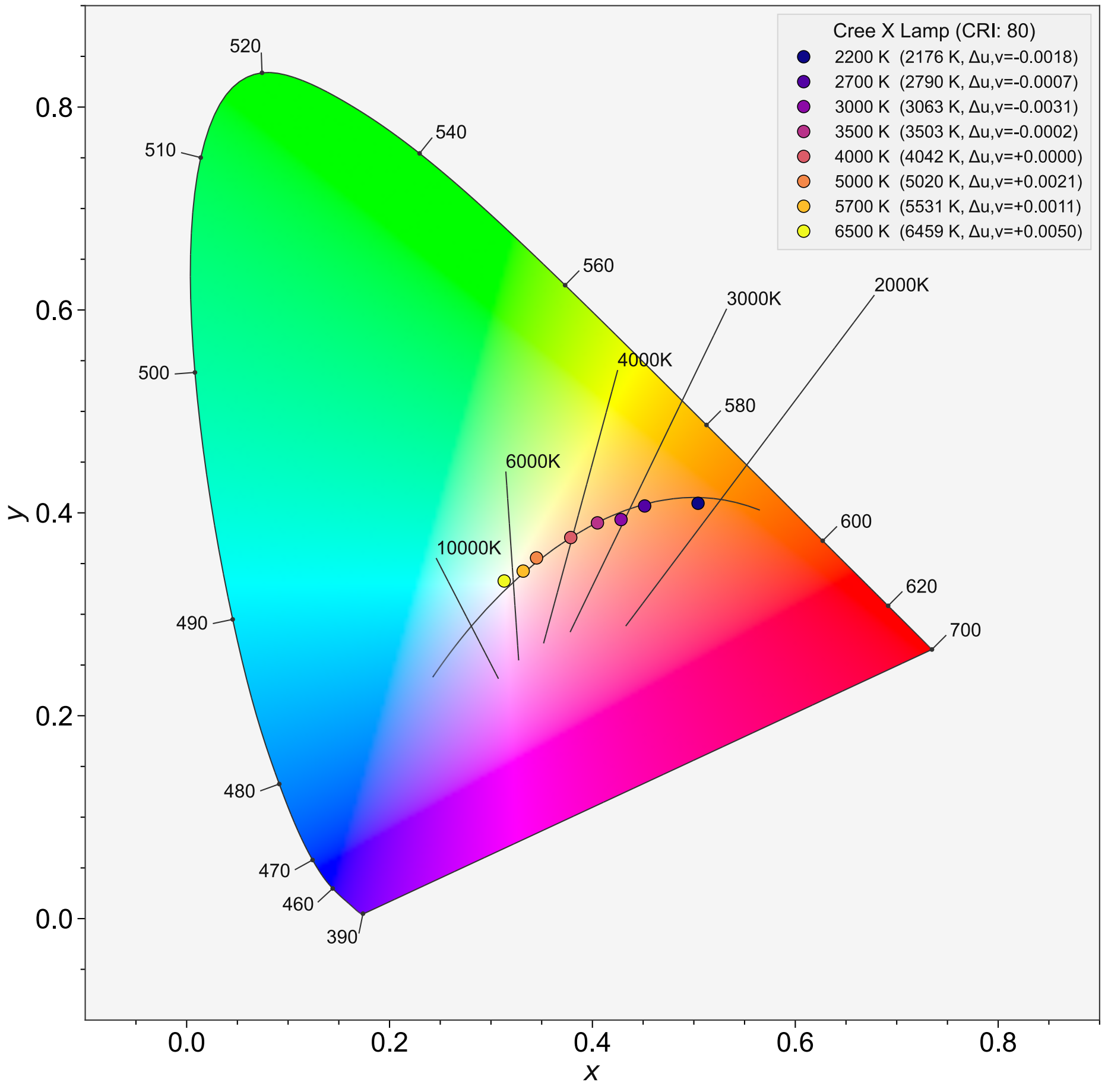

**Figure S16 –** Chromaticity chart for the 80 CRI Cree X Lamp LEDs.

**Table S5 -** Correlated color temperature values and chromaticity coordinates for the 80 CRI Cree XLamp LEDs.

| **Cree Xlamp (CRI: 80)** | | | | | | | |
|---|---|---|---|---|---|---|---|
| $CCT_0$ | CCT | ΔCCT | $x$ | $y$ | $u$ | $v$ | $\Delta uv$ |
| 2200 K | 2175.63 K | -24.37 K | 0.5041 | 0.4094 | 0.2920 | 0.3558 | -0.001840 |
| 2700 K | 2790.37 K | 90.37 K | 0.4515 | 0.4068 | 0.2588 | 0.3497 | -0.000672 |
| 3000 K | 3062.99 K | 62.99 K | 0.4282 | 0.3934 | 0.2495 | 0.3439 | -0.003103 |
| 3500 K | 3502.62 K | 2.62 K | 0.4050 | 0.3901 | 0.2357 | 0.3406 | -0.000186 |
| 4000 K | 4041.62 K | 41.62 K | 0.3786 | 0.3757 | 0.2244 | 0.3339 | 0.000030 |
| 5000 K | 5020.17 K | 20.17 K | 0.3449 | 0.3556 | 0.2098 | 0.3244 | 0.002064 |
| 5700 K | 5531.26 K | -168.74 K | 0.3317 | 0.3426 | 0.2058 | 0.3188 | 0.001110 |
| 6500 K | 6459.05 K | -40.95 K | 0.3130 | 0.3328 | 0.1966 | 0.3136 | 0.004994 |

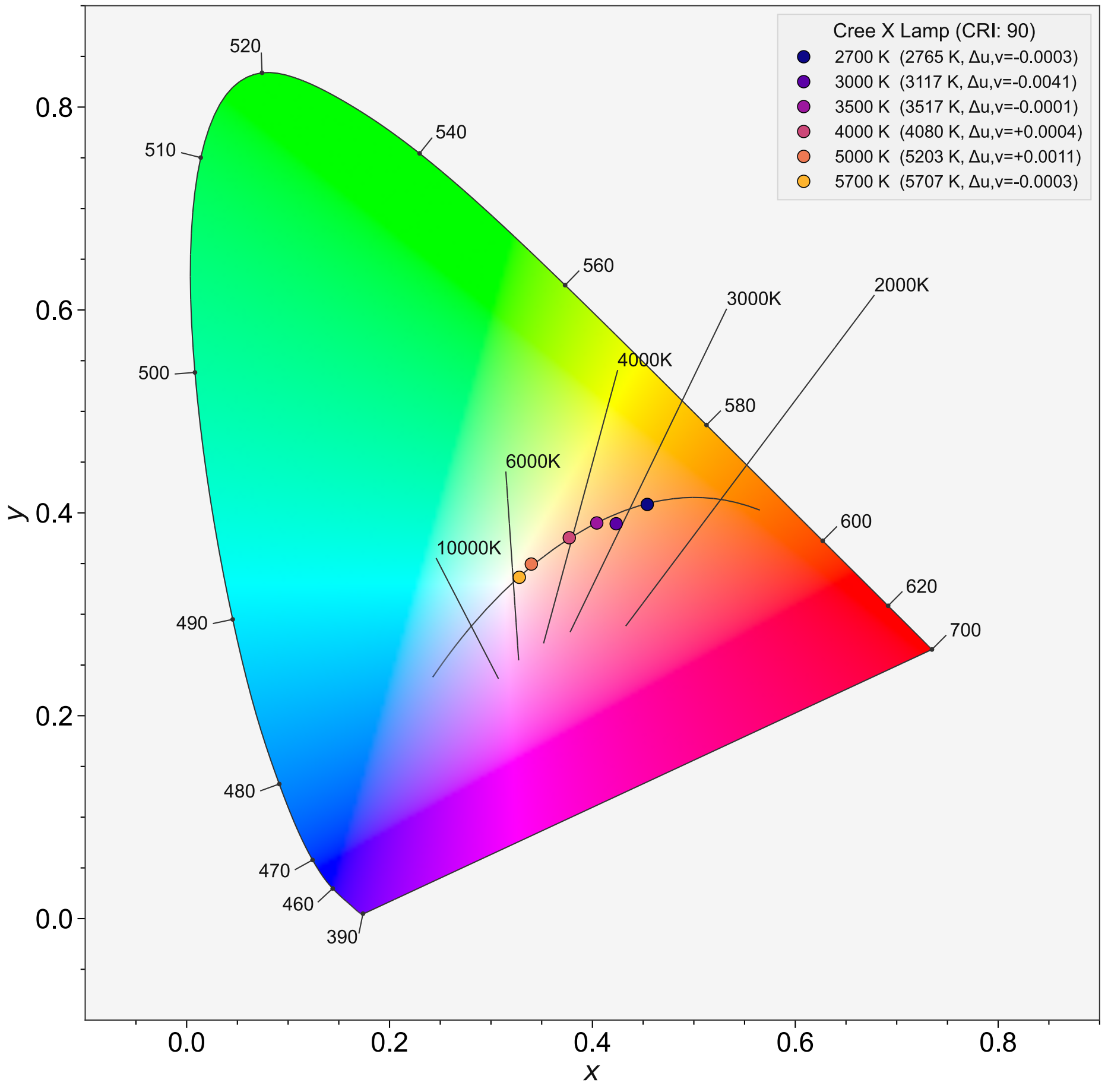

**Figure S17 –** Chromaticity chart for the 90 CRI Cree X Lamp LEDs.

**Table S6 -** Correlated color temperature values and chromaticity coordinates for the 90 CRI Cree XLamp LEDs.

| **Cree Xlamp (CRI: 90)** | | | | | | | |
|---|---|---|---|---|---|---|---|
| $CCT_0$ | CCT | ΔCCT | $x$ | $y$ | $u$ | $v$ | $\Delta uv$ |
| 2200 K | - | - | - | - | - | - | - |
| 2700 K | 2764.64 K | 64.64 K | 0.4541 | 0.4083 | 0.2598 | 0.3504 | -0.000341 |
| 3000 K | 3116.90 K | 116.90 K | 0.4234 | 0.3893 | 0.2482 | 0.3422 | -0.004078 |
| 3500 K | 3516.77 K | 16.77 K | 0.4043 | 0.3900 | 0.2353 | 0.3405 | -0.000079 |
| 4000 K | 4080.44 K | 80.44 K | 0.3772 | 0.3755 | 0.2235 | 0.3337 | 0.000359 |
| 5000 K | 5203.30 K | 203.30 K | 0.3398 | 0.3495 | 0.2086 | 0.3219 | 0.001148 |
| 5700 K | 5706.70 K | 6.70 K | 0.3279 | 0.3365 | 0.2055 | 0.3163 | -0.000300 |
| 6500 K | - | - | - | - | - | - | - |

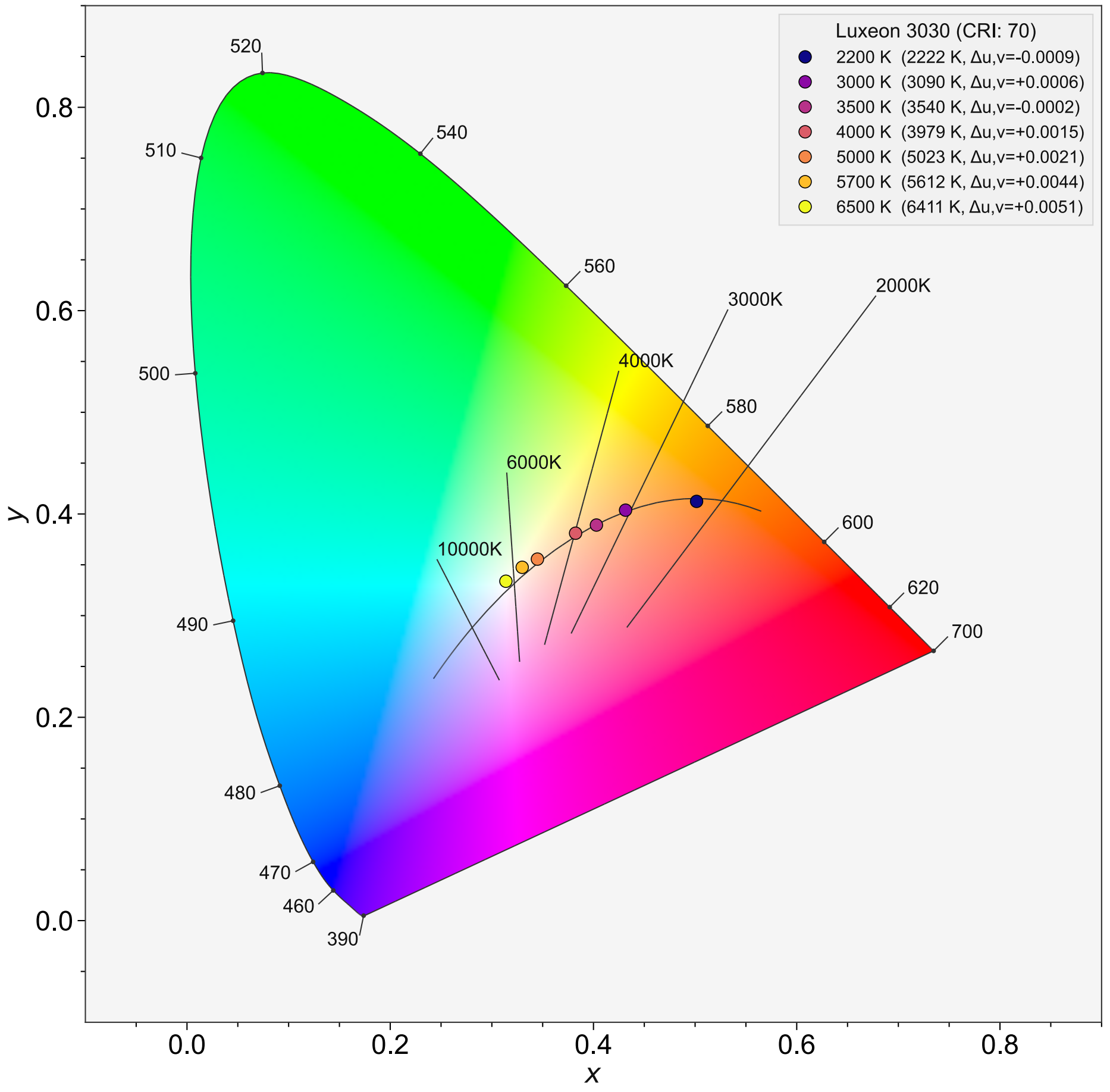

**Figure S18 –** Chromaticity chart for the 70 CRI Lumileds Luxeon 3030 LEDs.

**Table S7 -** Correlated color temperature values and chromaticity coordinates for the 70 CRI Lumileds Luxeon 3030 LEDs.

| **Luxeon 3030 (CRI: 70)** | | | | | | | |
|---|---|---|---|---|---|---|---|
| $CCT_0$ | CCT | ΔCCT | $x$ | $y$ | $u$ | $v$ | $\Delta uv$ |
| 2200 K | 2221.72 K | 21.72 K | 0.5015 | 0.4124 | 0.2888 | 0.3562 | -0.000921 |
| 2700 K | - | - | - | - | - | - | - |
| 3000 K | 3089.88 K | 89.88 K | 0.4316 | 0.4037 | 0.2472 | 0.3470 | 0.000642 |
| 3500 K | 3540.35 K | 40.35 K | 0.4029 | 0.3891 | 0.2348 | 0.3401 | -0.000177 |
| 4000 K | 3979.41 K | -20.59 K | 0.3824 | 0.3811 | 0.2247 | 0.3358 | 0.001454 |
| 5000 K | 5022.99 K | 22.99 K | 0.3448 | 0.3556 | 0.2097 | 0.3244 | 0.002100 |
| 5700 K | 5612.19 K | -87.81 K | 0.3298 | 0.3475 | 0.2027 | 0.3203 | 0.004440 |
| 6500 K | 6410.94 K | -89.06 K | 0.3138 | 0.3338 | 0.1968 | 0.3140 | 0.005111 |

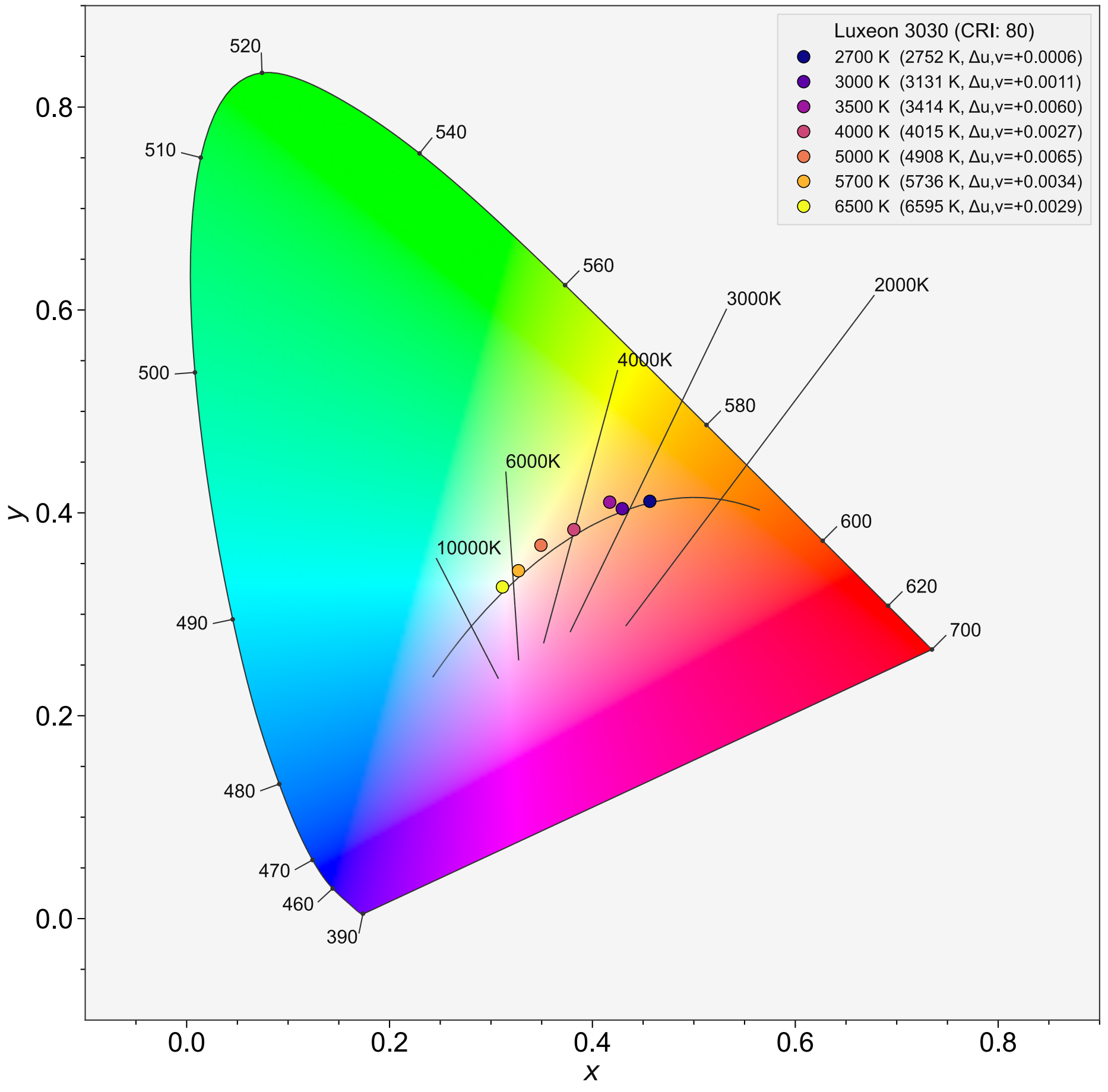

**Figure S19 –** Chromaticity chart for the 80 CRI Lumileds Luxeon 3030 LEDs.

**Table S8 -** Correlated color temperature values and chromaticity coordinates for the 80 CRI Lumileds Luxeon 3030 LEDs.

| **Luxeon 3030 (CRI: 80)** | | | | | | | |
|---|---|---|---|---|---|---|---|
| $CCT_0$ | CCT | ΔCCT | $x$ | $y$ | $u$ | $v$ | $\Delta uv$ |
| 2200 K | - | - | - | - | - | - | - |
| 2700 K | 2752.16 K | 52.16 K | 0.4566 | 0.4114 | 0.2600 | 0.3514 | 0.000580 |
| 3000 K | 3131.33 K | 131.33 K | 0.4294 | 0.4041 | 0.2457 | 0.3468 | 0.001117 |
| 3500 K | 3413.52 K | -86.48 K | 0.4171 | 0.4105 | 0.2352 | 0.3473 | 0.005961 |
| 4000 K | 4014.81 K | 14.81 K | 0.3817 | 0.3835 | 0.2233 | 0.3365 | 0.002749 |
| 5000 K | 4907.81 K | -92.19 K | 0.3493 | 0.3681 | 0.2079 | 0.3287 | 0.006472 |
| 5700 K | 5736.22 K | 36.22 K | 0.3271 | 0.3429 | 0.2025 | 0.3185 | 0.003382 |
| 6500 K | 6595.33 K | 95.33 K | 0.3113 | 0.3270 | 0.1976 | 0.3114 | 0.002868 |

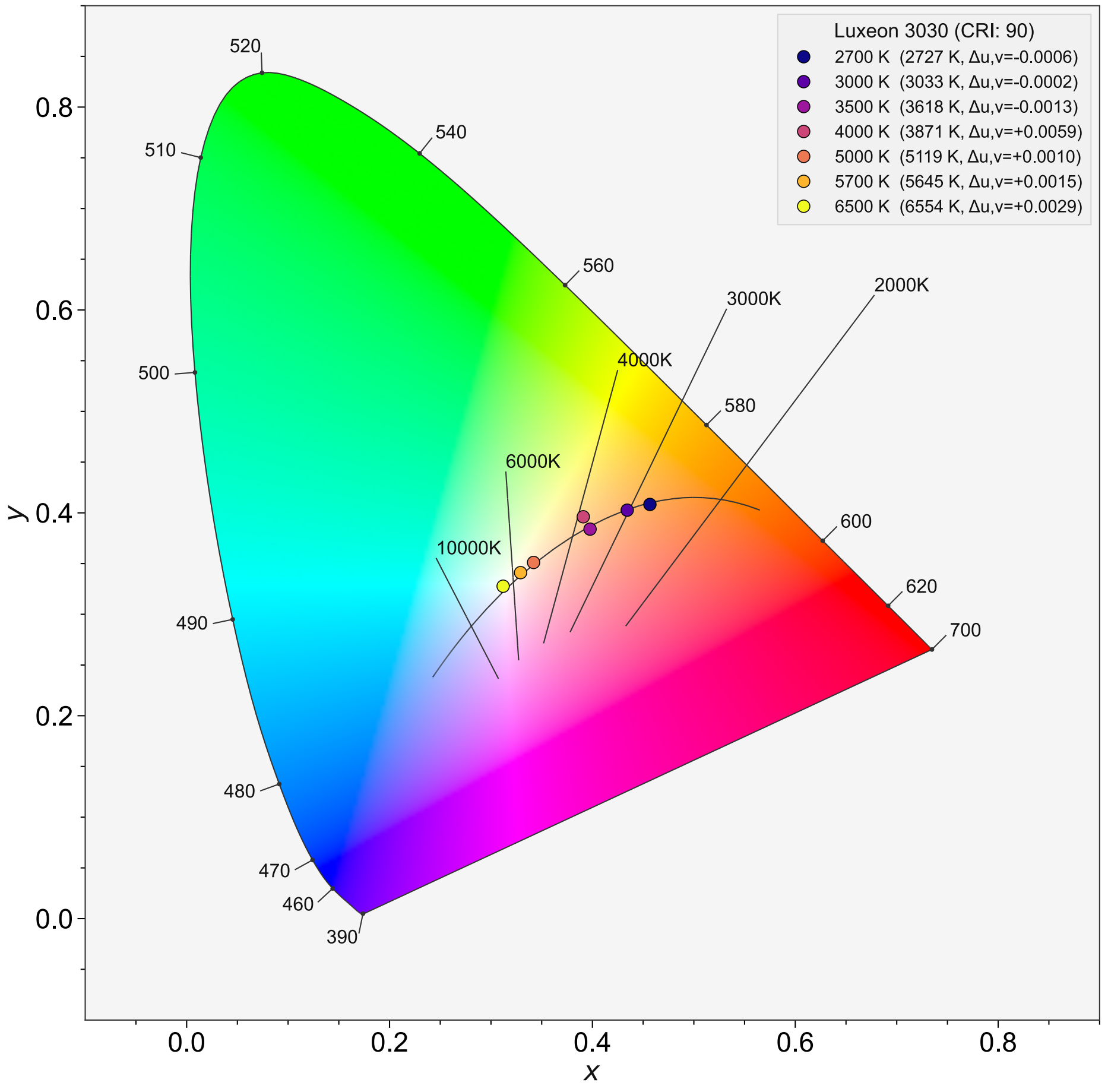

**Figure S20 –** Chromaticity chart for the 90 CRI Lumileds Luxeon 3030 LEDs.

**Table S9 -** Correlated color temperature values and chromaticity coordinates for the 90 CRI Lumileds Luxeon 3030 LEDs.

| **Luxeon 3030 (CRI: 90)** | | | | | | | |
|---|---|---|---|---|---|---|---|
| $CCT_0$ | CCT | ΔCCT | $x$ | $y$ | $u$ | $v$ | $\Delta uv$ |
| 2200 K | - | - | - | - | - | - | - |
| 2700 K | 2726.71 K | 26.71 K | 0.4567 | 0.4083 | 0.2615 | 0.3507 | -0.000587 |
| 3000 K | 3033.03 K | 33.03 K | 0.4343 | 0.4027 | 0.2495 | 0.3470 | -0.000203 |
| 3500 K | 3618.20 K | 118.20 K | 0.3977 | 0.3839 | 0.2335 | 0.3382 | -0.001264 |
| 4000 K | 3871.07 K | -128.93 K | 0.3910 | 0.3961 | 0.2243 | 0.3409 | 0.005860 |
| 5000 K | 5119.09 K | 119.09 K | 0.3420 | 0.3511 | 0.2095 | 0.3226 | 0.001012 |
| 5700 K | 5645.34 K | -54.66 K | 0.3292 | 0.3411 | 0.2046 | 0.3180 | 0.001482 |
| 6500 K | 6553.92 K | 53.92 K | 0.3120 | 0.3277 | 0.1978 | 0.3117 | 0.002897 |

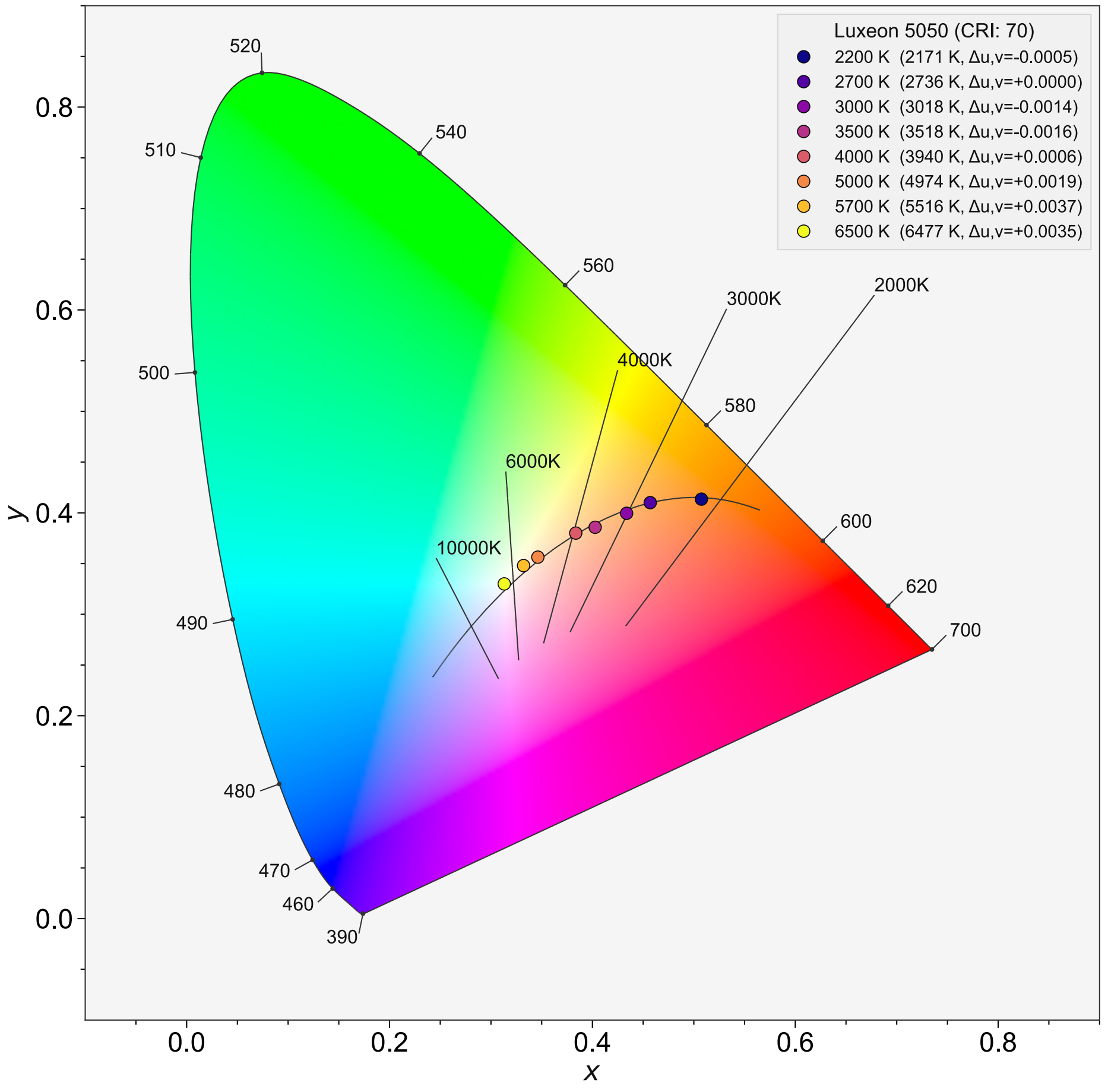

**Figure S21 –** Chromaticity chart for the 70 CRI Lumileds Luxeon 5050 LEDs.

**Table S10 -** Correlated color temperature values and chromaticity coordinates for the 70 CRI Lumileds Luxeon 5050 LEDs.

| **Luxeon 5050 (CRI: 70)** | | | | | | | |
|---|---|---|---|---|---|---|---|
| $CCT_0$ | CCT | ΔCCT | $x$ | $y$ | $u$ | $v$ | $\Delta uv$ |
| 2200 K | 2171.09 K | -28.91 K | 0.5075 | 0.4135 | 0.2922 | 0.3571 | -0.000512 |
| 2700 K | 2736.25 K | 36.25 K | 0.4570 | 0.4101 | 0.2609 | 0.3511 | 0.000049 |
| 3000 K | 3017.58 K | 17.58 K | 0.4337 | 0.3996 | 0.2504 | 0.3461 | -0.001366 |
| 3500 K | 3517.62 K | 17.62 K | 0.4027 | 0.3858 | 0.2360 | 0.3392 | -0.001604 |
| 4000 K | 3939.56 K | -60.44 K | 0.3836 | 0.3800 | 0.2259 | 0.3357 | 0.000625 |
| 5000 K | 4974.02 K | -25.98 K | 0.3462 | 0.3563 | 0.2104 | 0.3248 | 0.001919 |
| 5700 K | 5515.71 K | -184.29 K | 0.3321 | 0.3480 | 0.2040 | 0.3207 | 0.003678 |
| 6500 K | 6477.01 K | -22.99 K | 0.3131 | 0.3300 | 0.1977 | 0.3126 | 0.003521 |

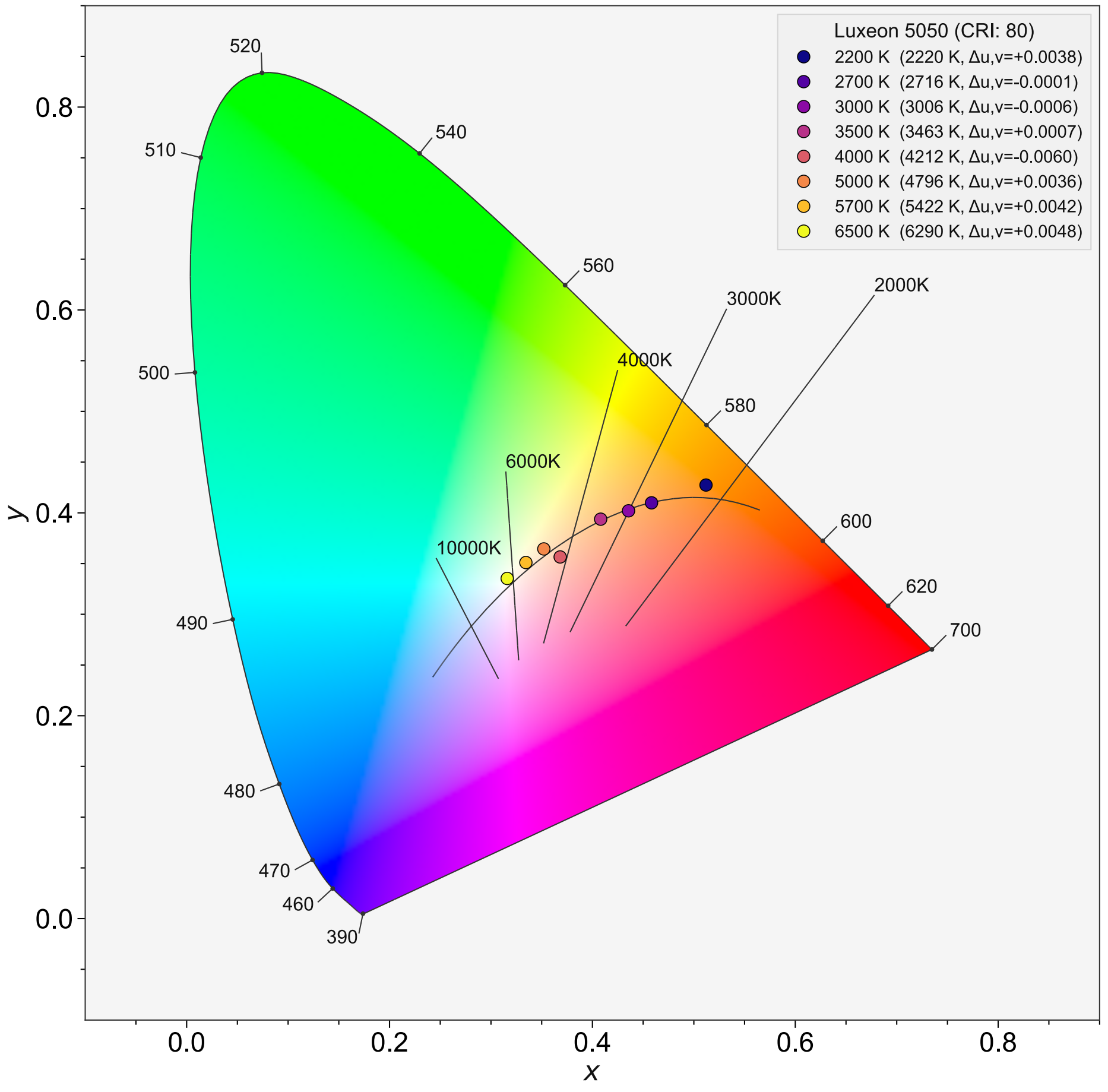

**Figure S22 –** Chromaticity chart for the 80 CRI Lumileds Luxeon 5050 LEDs.

**Table S11 -** Correlated color temperature values and chromaticity coordinates for the 80 CRI Lumileds Luxeon 5050 LEDs.

| **Luxeon 5050 (CRI: 80)** | | | | | | | |
|---|---|---|---|---|---|---|---|
| $CCT_0$ | CCT | ΔCCT | $x$ | $y$ | $u$ | $v$ | $\Delta uv$ |
| 2200 K | 2219.64 K | 19.64 K | 0.5120 | 0.4274 | 0.2882 | 0.3609 | 0.003814 |
| 2700 K | 2716.43 K | 16.43 K | 0.4583 | 0.4099 | 0.2618 | 0.3512 | -0.000132 |
| 3000 K | 3005.53 K | 5.53 K | 0.4356 | 0.4020 | 0.2506 | 0.3469 | -0.000641 |
| 3500 K | 3463.44 K | -36.56 K | 0.4081 | 0.3938 | 0.2363 | 0.3420 | 0.000718 |
| 4000 K | 4212.24 K | 212.24 K | 0.3682 | 0.3565 | 0.2251 | 0.3270 | -0.006015 |
| 5000 K | 4796.33 K | -203.67 K | 0.3521 | 0.3644 | 0.2112 | 0.3279 | 0.003610 |
| 5700 K | 5421.95 K | -278.05 K | 0.3344 | 0.3510 | 0.2044 | 0.3219 | 0.004177 |
| 6500 K | 6289.91 K | -210.09 K | 0.3160 | 0.3353 | 0.1978 | 0.3148 | 0.004789 |

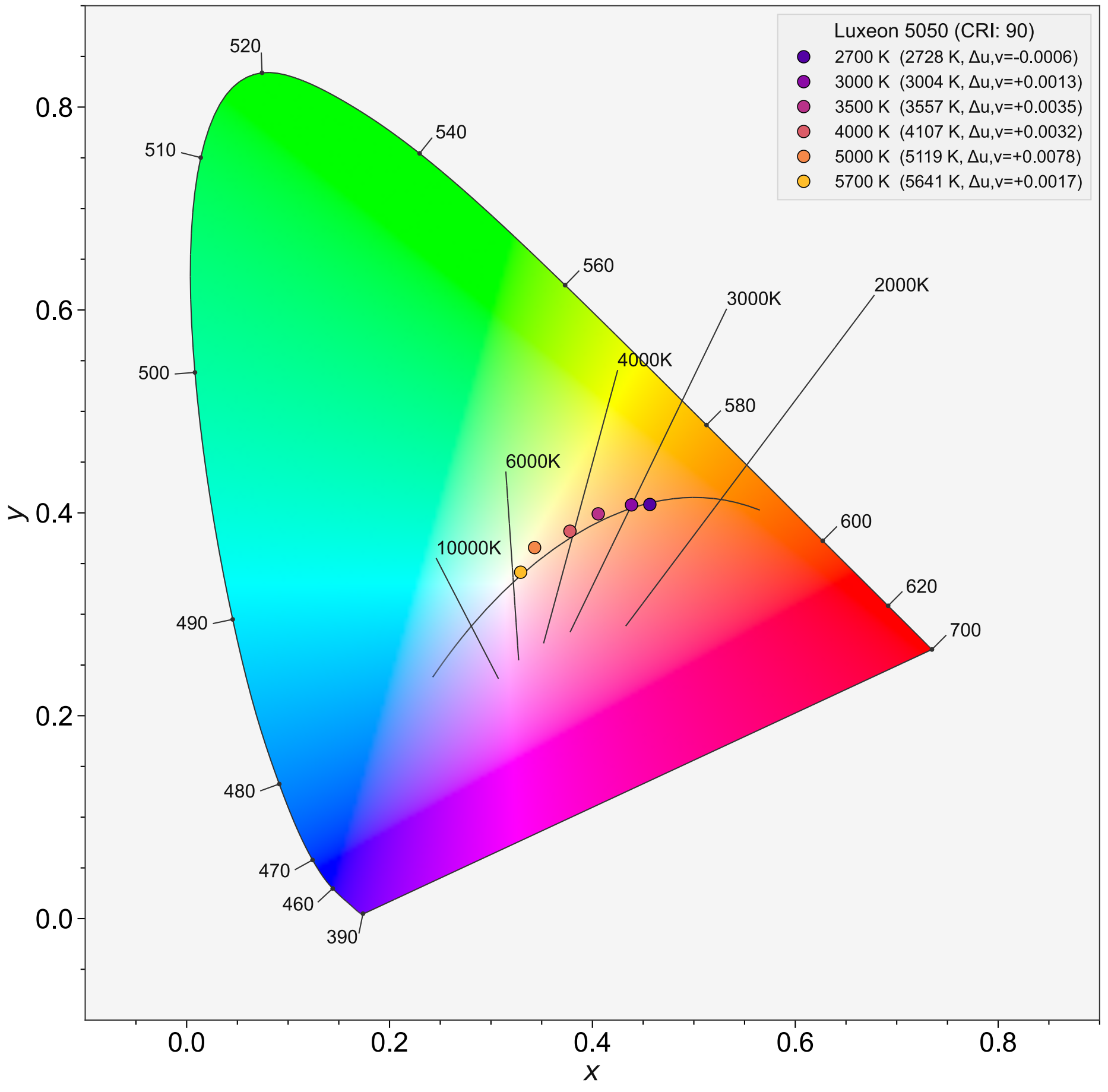

**Figure S23 –** Chromaticity chart for the 90 CRI Lumileds Luxeon 5050 LEDs.

**Table S12 -** Correlated color temperature values and chromaticity coordinates for the 90 CRI Lumileds Luxeon 5050 LEDs.

| **Luxeon 5050 (CRI: 90)** | | | | | | | |
|---|---|---|---|---|---|---|---|
| $CCT_0$ | CCT | ΔCCT | $x$ | $y$ | $u$ | $v$ | $\Delta uv$ |
| 2200 K | - | - | - | - | - | - | - |
| 2700 K | 2727.53 K | 27.53 K | 0.4566 | 0.4083 | 0.2615 | 0.3506 | -0.000589 |
| 3000 K | 3004.06 K | 4.06 K | 0.4385 | 0.4079 | 0.2500 | 0.3487 | 0.001283 |
| 3500 K | 3557.08 K | 57.08 K | 0.4057 | 0.3990 | 0.2326 | 0.3432 | 0.003506 |
| 4000 K | 4107.10 K | 107.10 K | 0.3779 | 0.3819 | 0.2214 | 0.3356 | 0.003165 |
| 5000 K | 5119.42 K | 119.42 K | 0.3430 | 0.3658 | 0.2047 | 0.3274 | 0.007798 |
| 5700 K | 5641.47 K | -58.53 K | 0.3293 | 0.3415 | 0.2045 | 0.3182 | 0.001657 |
| 6500 K | - | - | - | - | - | - | - |

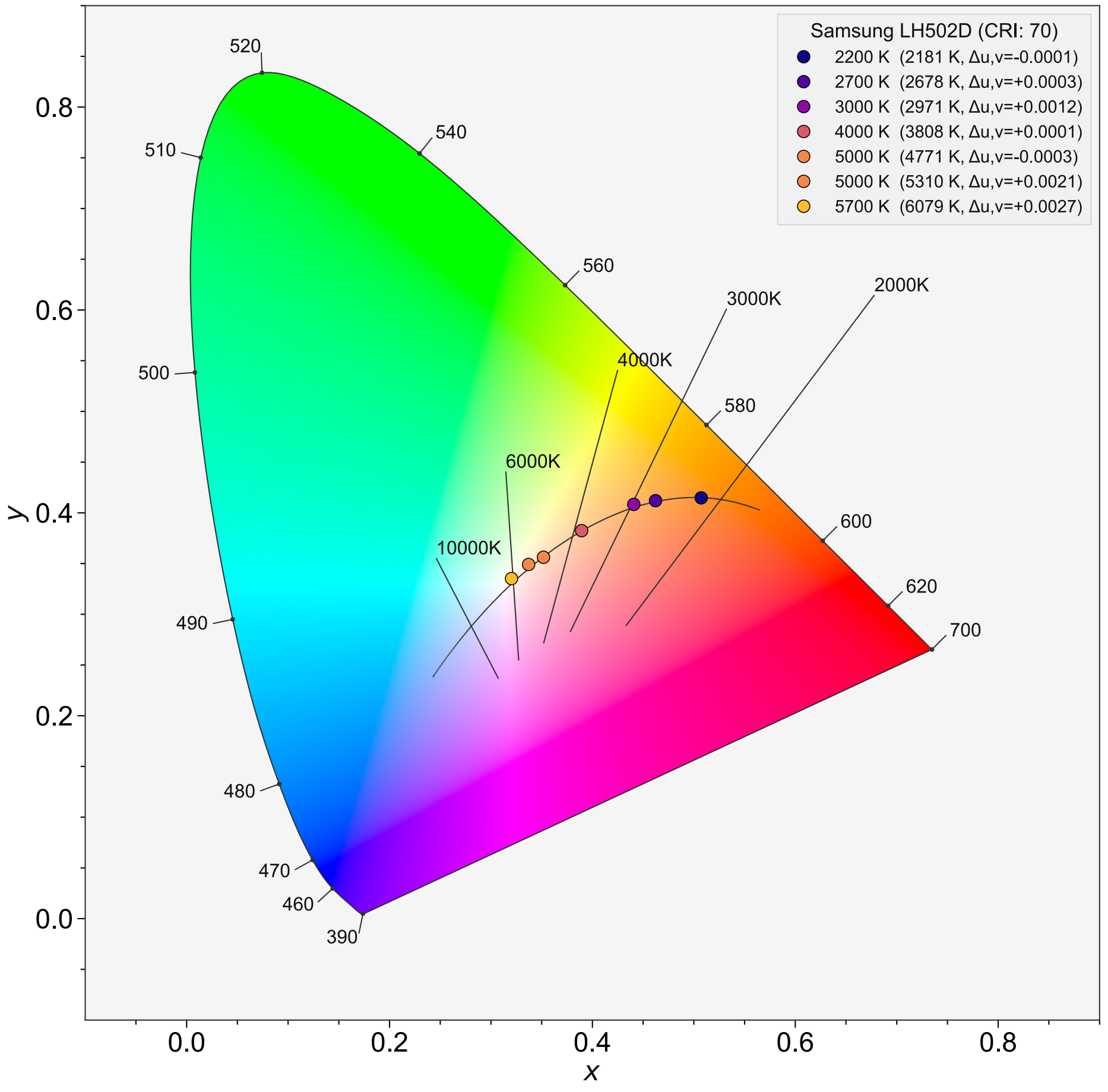

**Figure S24 –** Chromaticity chart for the 70 CRI Samsung LH502D LEDs.

**Table 13 -** Correlated color temperature values and chromaticity coordinates for the 70 CRI Samsung LH502D LEDs.

| **Samsung LH502D (CRI: 70)** | | | | | | | |
|---|---|---|---|---|---|---|---|
| $CCT_0$ | CCT | ΔCCT | $x$ | $y$ | $u$ | $v$ | $\Delta uv$ |
| 2200 K | 2181.41 K | -18.59 K | 0.5073 | 0.4148 | 0.2914 | 0.3574 | -0.000118 |
| 2700 K | 2678.29 K | -21.71 K | 0.4622 | 0.4120 | 0.2634 | 0.3522 | 0.000333 |
| 3000 K | 2971.02 K | -28.99 K | 0.4408 | 0.4083 | 0.2512 | 0.3491 | 0.001172 |
| 3500 K | - | - | - | - | - | - | - |
| 4000 K | 3808.43 K | -191.98 K | 0.3894 | 0.3824 | 0.2287 | 0.3369 | 0.000144 |
| 5000 K | 4770.94 K | -229.06 K | 0.3518 | 0.3562 | 0.2141 | 0.3253 | -0.000312 |
| 5700 K | 5310.19 K | -389.81 K | 0.3371 | 0.3491 | 0.2070 | 0.3215 | 0.002077 |
| 6500 K | 6078.52 K | -421.48 K | 0.3203 | 0.3352 | 0.2007 | 0.3151 | 0.002659 |

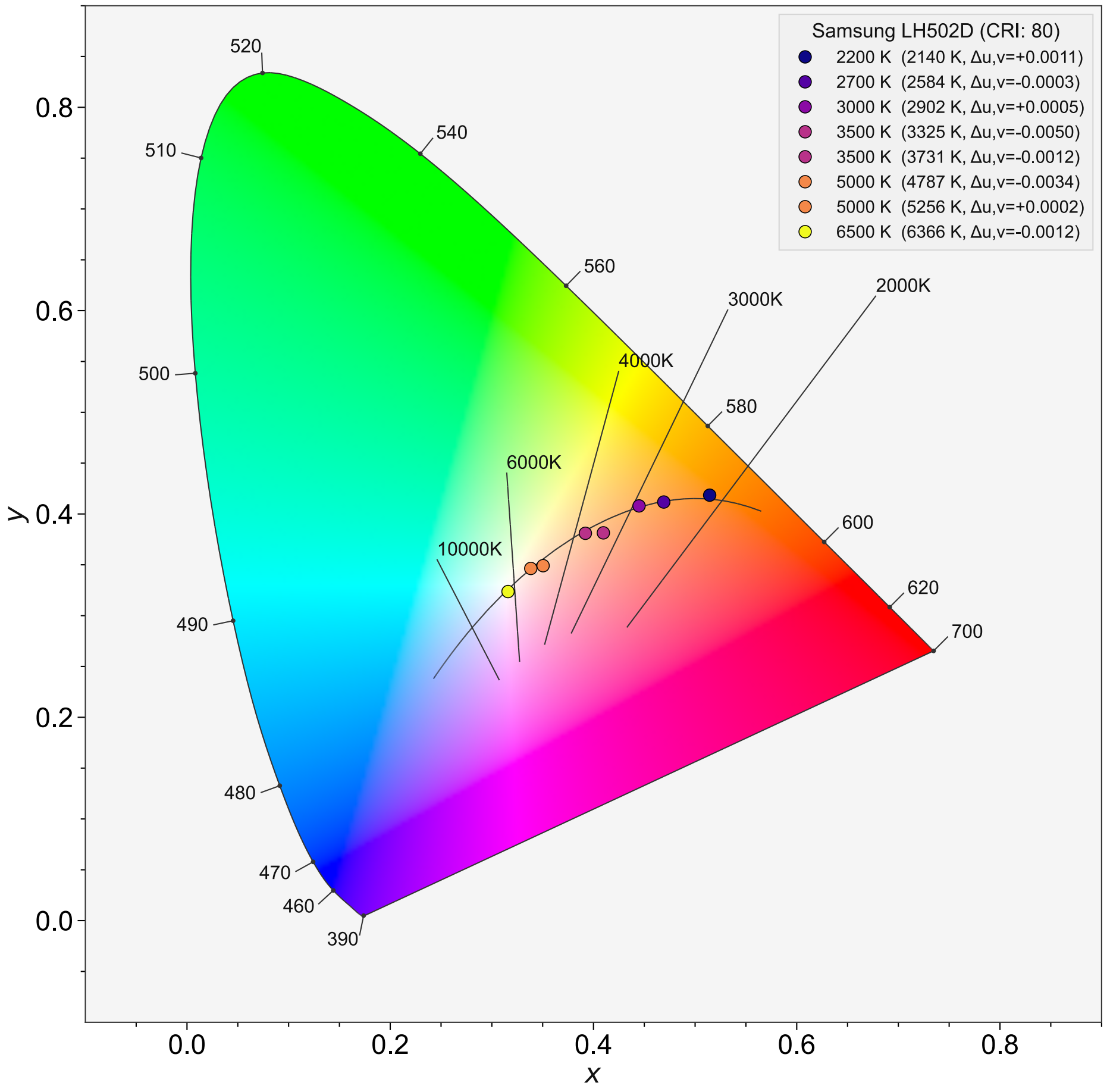

**Figure S25 –** Chromaticity chart for the 80 CRI Samsung LH502D LEDs.

**Table S14 -** Correlated color temperature values and chromaticity coordinates for the 80 CRI Samsung LH502D LEDs.

| **Samsung LH502D (CRI: 80)** | | | | | | | |
|---|---|---|---|---|---|---|---|
| $CCT_0$ | CCT | ΔCCT | $x$ | $y$ | $u$ | $v$ | $\Delta uv$ |
| 2200 K | 2139.72 K | -60.28 K | 0.5143 | 0.4185 | 0.2942 | 0.3591 | 0.001140 |
| 2700 K | 2583.53 K | -116.47 K | 0.4692 | 0.4117 | 0.2680 | 0.3528 | -0.000260 |
| 3000 K | 2902.30 K | -97.70 K | 0.4448 | 0.4079 | 0.2540 | 0.3494 | 0.000503 |
| 3500 K | 3324.61 K | -175.39 K | 0.4097 | 0.3814 | 0.2425 | 0.3387 | -0.005034 |
| 4000 K | 3731.12 K | -268.88 K | 0.3920 | 0.3810 | 0.2310 | 0.3368 | -0.001204 |
| 5000 K | 4787.40 K | -212.60 K | 0.3504 | 0.3491 | 0.2160 | 0.3228 | -0.003372 |
| 5700 K | 5256.41 K | -443.59 K | 0.3383 | 0.3464 | 0.2088 | 0.3207 | 0.000194 |
| 6500 K | 6365.92 K | -134.08 K | 0.3159 | 0.3237 | 0.2021 | 0.3106 | -0.001190 |

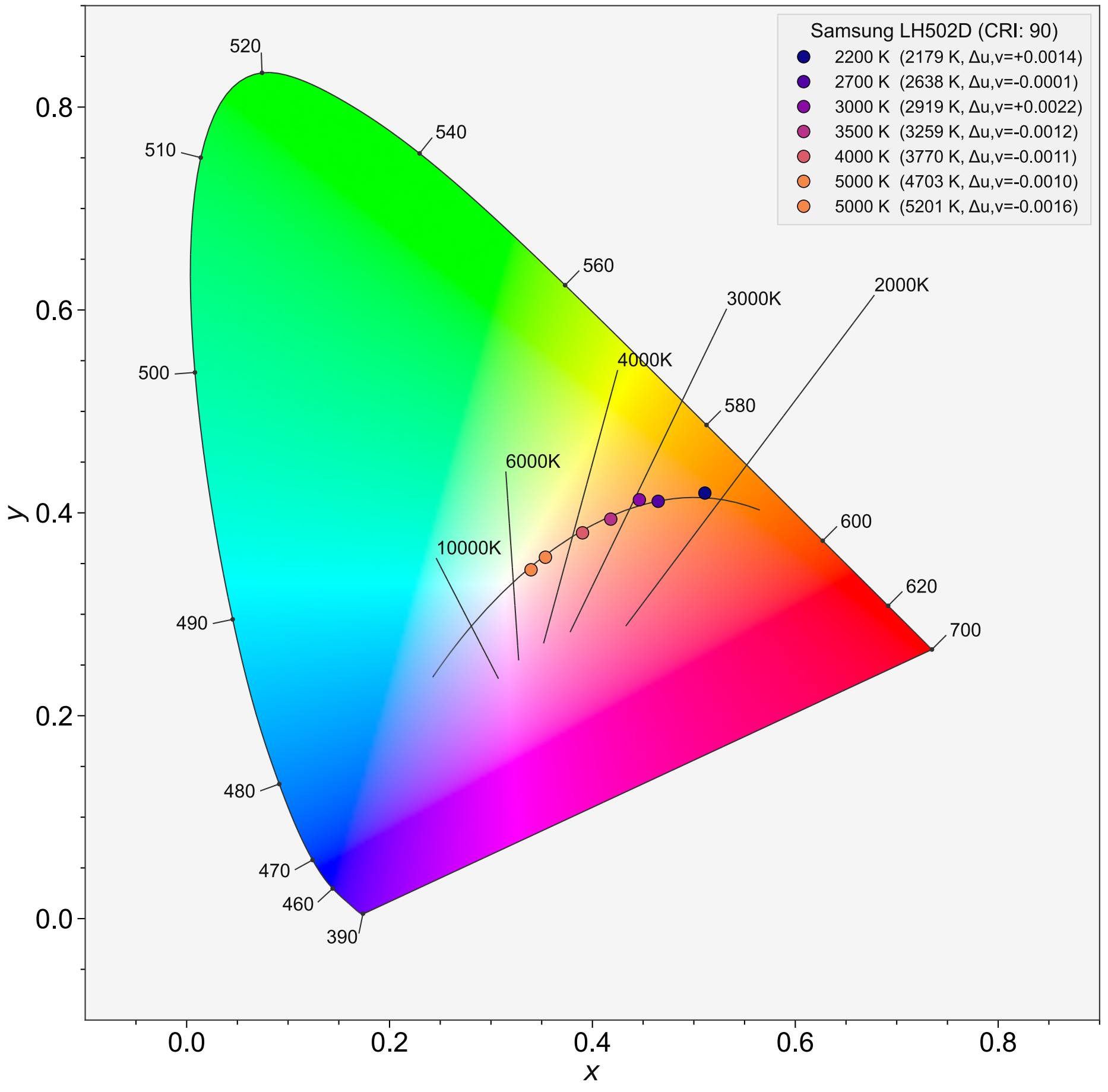

**Figure S26 –** Chromaticity chart for the 90 CRI Samsung LH502D LEDs.

**Table S15 -** Correlated color temperature values and chromaticity coordinates for the 90 CRI Samsung LH502D LEDs.

| **Samsung LH502D (CRI: 90)** | | | | | | | |
|---|---|---|---|---|---|---|---|
| $CCT_0$ | CCT | ΔCCT | $x$ | $y$ | $u$ | $v$ | $\Delta uv$ |
| 2200 K | 2178.65 K | -21.35 K | 0.5109 | 0.4195 | 0.2914 | 0.3589 | 0.001393 |
| 2700 K | 2637.94 K | -62.06 K | 0.4648 | 0.4114 | 0.2654 | 0.3523 | -0.000097 |
| 3000 K | 2918.54 K | -81.46 K | 0.4463 | 0.4129 | 0.2528 | 0.3508 | 0.002236 |
| 3500 K | 3258.94 K | -241.06 K | 0.4181 | 0.3938 | 0.2427 | 0.3430 | -0.001226 |
| 4000 K | 3770.02 K | -229.98 K | 0.3902 | 0.3803 | 0.2301 | 0.3364 | -0.001064 |
| 5000 K | 4702.84 K | -297.16 K | 0.3537 | 0.3562 | 0.2154 | 0.3255 | -0.001023 |
| 5700 K | 5201.06 K | -498.94 K | 0.3395 | 0.3439 | 0.2106 | 0.3200 | -0.001615 |
| 6500 K | - | - | - | - | - | - | - |

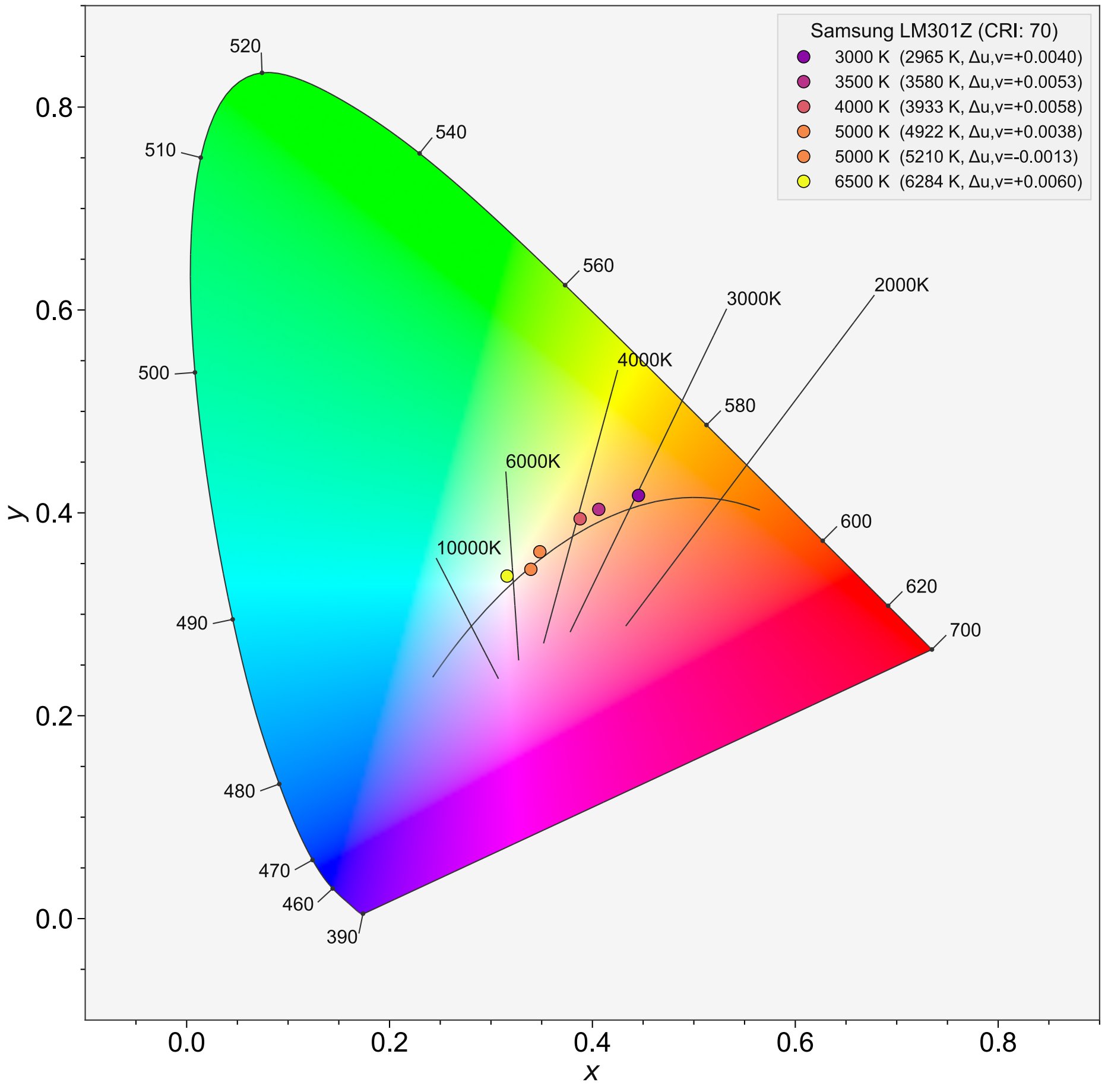

**Figure S27 –** Chromaticity chart for the 70 CRI Samsung LM301Z LEDs.

**Table S16 -** Correlated color temperature values and chromaticity coordinates for the 70 CRI Samsung LM301Z LEDs.

| **Samsung LM301Z (CRI: 70)** | | | | | | | |
|---|---|---|---|---|---|---|---|
| $CCT_0$ | CCT | ΔCCT | $x$ | $y$ | $u$ | $v$ | $\Delta uv$ |
| 2200 K | - | - | - | - | - | - | - |
| 2700 K | - | - | - | - | - | - | - |
| 3000 K | 2965.20 K | -34.80 K | 0.4454 | 0.4171 | 0.2504 | 0.3518 | 0.003967 |
| 3500 K | 3579.64 K | 79.64 K | 0.4063 | 0.4034 | 0.2312 | 0.3444 | 0.005277 |
| 4000 K | 3932.67 K | -67.33 K | 0.3879 | 0.3941 | 0.2231 | 0.3401 | 0.005830 |
| 5000 K | 4922.08 K | -77.92 K | 0.3482 | 0.3616 | 0.2096 | 0.3266 | 0.003758 |
| 5700 K | 5209.81 K | -490.19 K | 0.3393 | 0.3443 | 0.2103 | 0.3201 | -0.001301 |
| 6500 K | 6283.72 K | -216.28 K | 0.3159 | 0.3377 | 0.1968 | 0.3156 | 0.006044 |

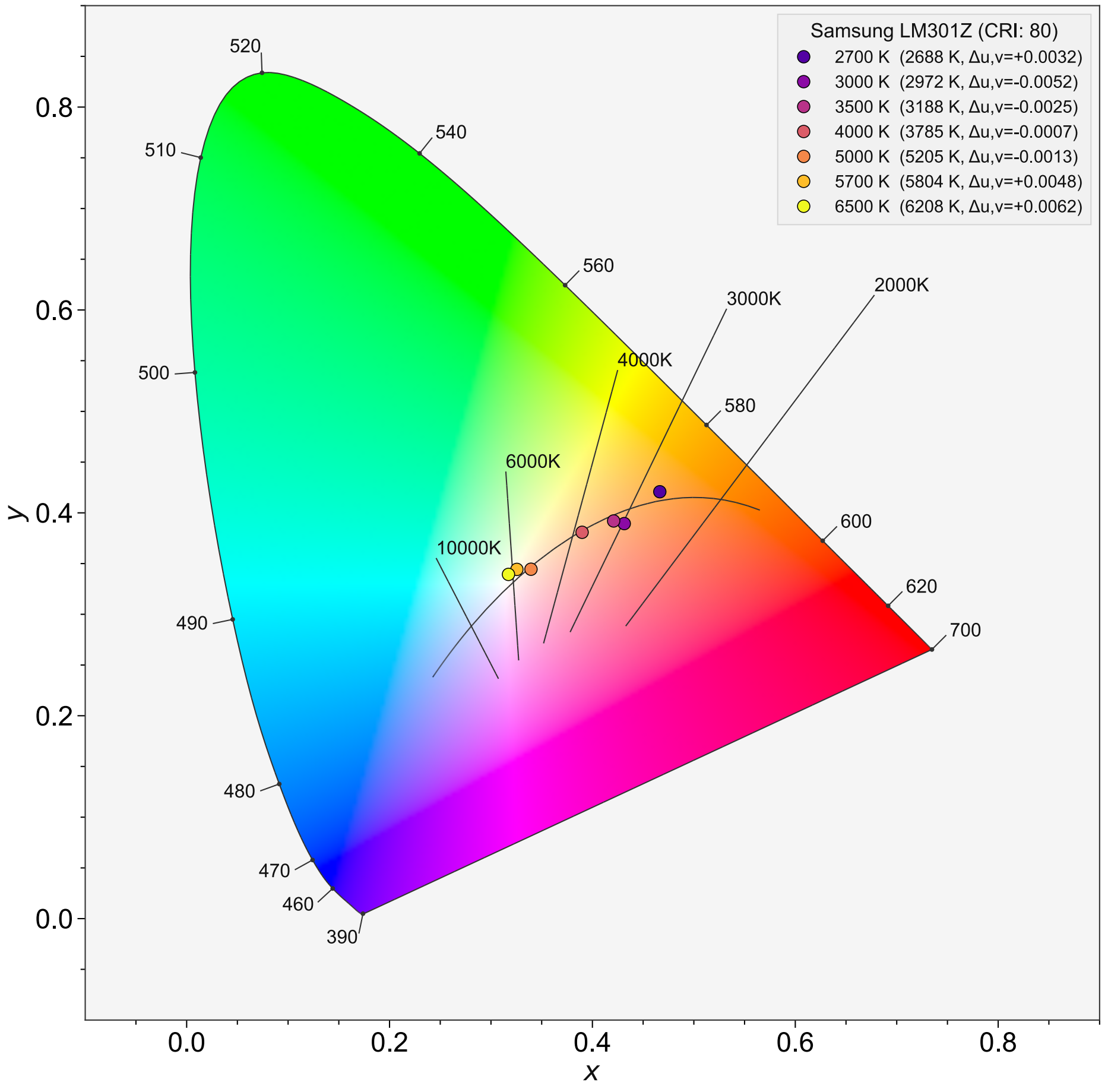

**Figure S28 –** Chromaticity chart for the 80 CRI Samsung LM301Z LEDs.

**Table S17 -** Correlated color temperature values and chromaticity coordinates for the 80 CRI Samsung LM301Z LEDs.

| **Samsung LM301Z (CRI: 80)** | | | | | | | |
|---|---|---|---|---|---|---|---|
| $CCT_0$ | CCT | ΔCCT | $x$ | $y$ | $u$ | $v$ | $\Delta uv$ |
| 2200 K | - | - | - | - | - | - | - |
| 2700 K | 2687.98 K | -12.02 K | 0.4664 | 0.4208 | 0.2622 | 0.3548 | 0.003171 |
| 3000 K | 2972.17 K | -27.83 K | 0.4314 | 0.3894 | 0.2534 | 0.3431 | -0.005208 |
| 3500 K | 3188.09 K | -311.91 K | 0.4209 | 0.3920 | 0.2454 | 0.3428 | -0.002504 |
| 4000 K | 3784.51 K | -215.49 K | 0.3898 | 0.3808 | 0.2296 | 0.3365 | -0.000709 |
| 5000 K | 5204.80 K | 204.80 K | 0.3394 | 0.3444 | 0.2104 | 0.3202 | -0.001309 |
| 5700 K | 5804.43 K | 104.43 K | 0.3255 | 0.3443 | 0.2009 | 0.3188 | 0.004759 |
| 6500 K | 6207.98 K | -292.02 K | 0.3172 | 0.3393 | 0.1971 | 0.3163 | 0.006200 |

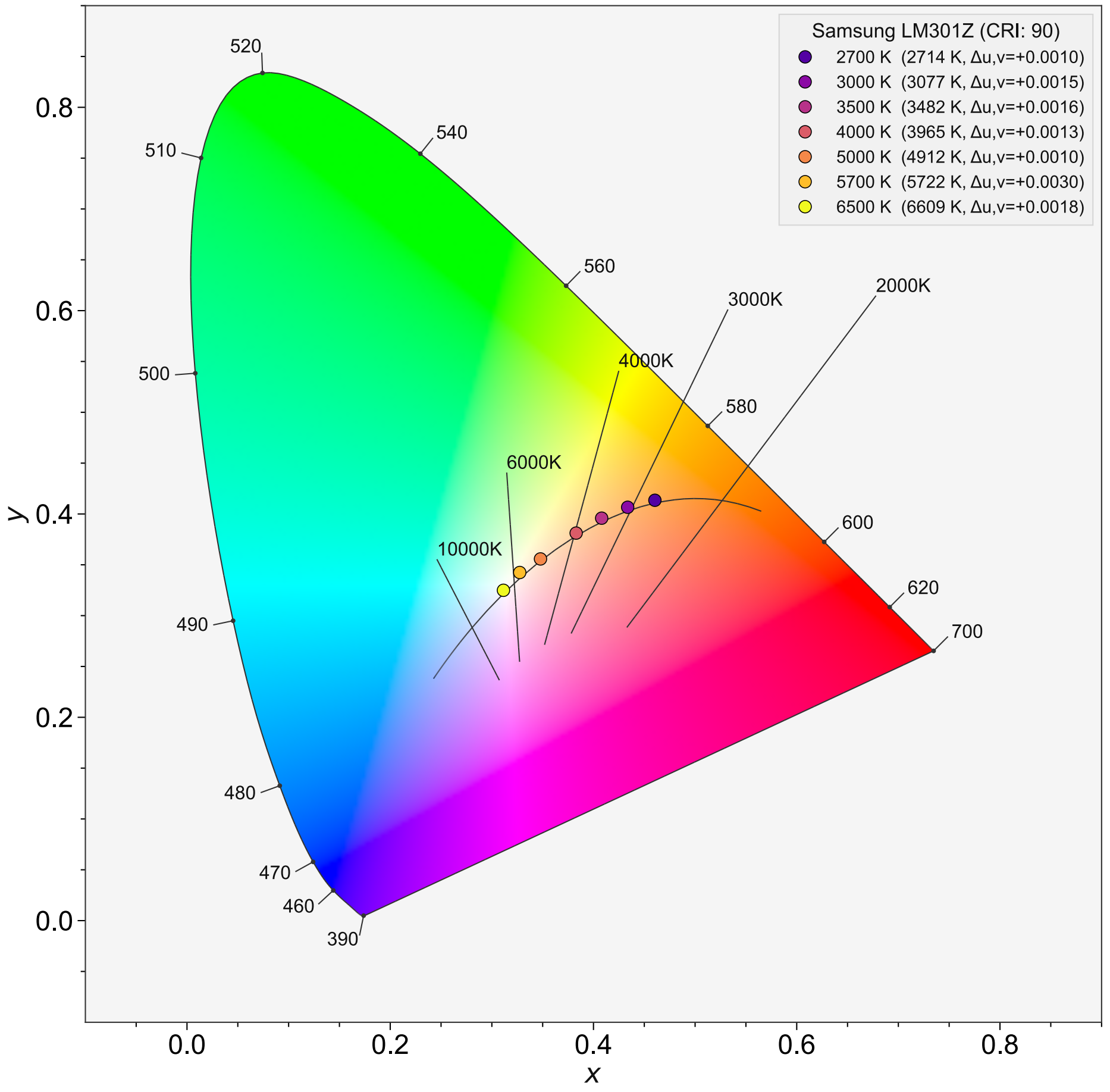

**Figure S29 –** Chromaticity chart for the 90 CRI Samsung LM301Z LEDs.

**Table S18 -** Correlated color temperature values and chromaticity coordinates for the 90 CRI Samsung LM301Z LEDs.

| **Samsung LM301Z (CRI: 90)** | | | | | | | |
|---|---|---|---|---|---|---|---|
| $CCT_0$ | CCT | ΔCCT | $x$ | $y$ | $u$ | $v$ | $\Delta uv$ |
| 2200 K | - | - | - | - | - | - | - |
| 2700 K | 2714.31 K | 14.31 K | 0.4604 | 0.4135 | 0.2616 | 0.3524 | 0.001012 |
| 3000 K | 3076.83 K | 76.83 K | 0.4336 | 0.4066 | 0.2474 | 0.3479 | 0.001479 |
| 3500 K | 3482.30 K | -17.70 K | 0.4081 | 0.3959 | 0.2354 | 0.3425 | 0.001644 |
| 4000 K | 3965.47 K | -34.53 K | 0.3829 | 0.3812 | 0.2250 | 0.3359 | 0.001345 |
| 5000 K | 4911.82 K | -88.18 K | 0.3479 | 0.3558 | 0.2117 | 0.3247 | 0.000988 |
| 5700 K | 5722.33 K | 22.33 K | 0.3274 | 0.3424 | 0.2029 | 0.3183 | 0.002969 |
| 6500 K | 6609.42 K | 109.42 K | 0.3114 | 0.3249 | 0.1985 | 0.3106 | 0.001767 |